\documentclass[a4paper,11pt]{article}
\usepackage{jinstpub} 

\usepackage{lineno}
\input{preamble}

\title{\boldmath Neutron detector response modeling in NOvA}

\collaboration{The NOvA Collaboration}

\makeatletter
\ExplSyntaxOn
\cs_new_eq:NN \nova_jinst_author:w \author
\cs_new_eq:NN \nova_jinst_affiliation:w \affiliation
\prop_new:N \g_nova_affiliation_numbers_prop
\seq_new:N \g_nova_current_author_affiliations_seq
\int_new:N \g_nova_affiliation_int
\tl_new:N \g_nova_pending_author_tl
\bool_new:N \g_nova_seen_author_bool
\cs_new_protected:Npn \nova_emit_author:nn #1#2
  { \nova_jinst_author:w[#1]{#2} }
\cs_generate_variant:Nn \nova_emit_author:nn { xV }
\cs_new_protected:Npn \nova_emit_affiliation:nn #1#2
  { \nova_jinst_affiliation:w[#1]{#2} }
\cs_generate_variant:Nn \nova_emit_affiliation:nn { xn }
\cs_new_protected:Npn \nova_flush_pending_author:
  {
    \tl_if_empty:NF \g_nova_pending_author_tl
      {
        \nova_emit_author:xV
          { \seq_use:Nn \g_nova_current_author_affiliations_seq {,} }
          \g_nova_pending_author_tl
        \tl_gclear:N \g_nova_pending_author_tl
        \seq_gclear:N \g_nova_current_author_affiliations_seq
      }
  }
\RenewDocumentCommand{\author}{O{}m}
  {
    \nova_flush_pending_author:
    \bool_gset_true:N \g_nova_seen_author_bool
    \tl_gset:Nn \g_nova_pending_author_tl {#2}
    \seq_gclear:N \g_nova_current_author_affiliations_seq
    \tl_if_blank:nF {#1}
      { \seq_gset_split:Nnn \g_nova_current_author_affiliations_seq {,} {#1} }
  }
\RenewDocumentCommand{\affiliation}{O{}m}
  {
    \tl_set:Nx \l_tmpa_tl { \tl_to_str:n {#2} }
    \bool_if:NTF \g_nova_seen_author_bool
      {
        \prop_get:NVNTF \g_nova_affiliation_numbers_prop \l_tmpa_tl \l_tmpb_tl
          { \seq_gput_right:NV \g_nova_current_author_affiliations_seq \l_tmpb_tl }
          { \PackageError{nova-authors}{Author affiliation not found in affiliation list:~\tl_to_str:n {#2}}{} }
      }
      {
        \tl_if_blank:nTF {#1}
          { \int_gincr:N \g_nova_affiliation_int }
          { \int_gset:Nn \g_nova_affiliation_int {#1} }
        \prop_gput:NVx \g_nova_affiliation_numbers_prop \l_tmpa_tl { \int_use:N \g_nova_affiliation_int }
        \nova_emit_affiliation:xn { \int_use:N \g_nova_affiliation_int } {#2}
      }
  }
\ExplSyntaxOff
\makeatother
\newcommand{\ANL}{Argonne National Laboratory, Argonne, Illinois 60439, 
USA}
\newcommand{\Bandirma}{Bandirma Onyedi Eyl\"ul University, Faculty of 
Engineering and Natural Sciences, Engineering Sciences Department, 
10200, Bandirma, Balıkesir, Turkey}
\newcommand{\ICS}{Institute of Computer Science, The Czech 
Academy of Sciences, 
182 07 Prague, Czech Republic}
\newcommand{\IOP}{Institute of Physics, The Czech 
Academy of Sciences, 
182 21 Prague, Czech Republic}
\newcommand{\Atlantico}{Universidad del Atlantico,
Carrera 30 No.\ 8-49, Puerto Colombia, Atlantico, Colombia}
\newcommand{\BHU}{Department of Physics, Institute of Science, Banaras 
Hindu University, Varanasi, 221 005, India}

\newcommand{\Caltech}{California Institute of 
Technology, Pasadena, California 91125, USA}
\newcommand{\CUSB}{Central University of South Bihar,
NH-120, Gaya Panchanpur Road, Post Fatehpur, 
Gaya, Bihar, 824 236, India}
\newcommand{\Cochin}{Department of Physics, Cochin University
of Science and Technology, Kochi 682 022, India}
\newcommand{\Charles}
{Charles University, Faculty of Mathematics and Physics,
 Institute of Particle and Nuclear Physics, Prague, Czech Republic}
\newcommand{\Cincinnati}{Department of Physics, University of Cincinnati, 
Cincinnati, Ohio 45221, USA}
\newcommand{\CSU}{Department of Physics, Colorado 
State University, Fort Collins, CO 80523-1875, USA}
\newcommand{\CTU}{Czech Technical University in Prague,
Brehova 7, 115 19 Prague 1, Czech Republic}

\newcommand{\Delhi}{Department of Physics and Astrophysics, University of 
Delhi, Delhi 110007, India}
\newcommand{\JINR}{Joint Institute for Nuclear Research,  
Dubna, Moscow region 141980, Russia}
\newcommand{\Erciyes}{
Department of Physics, Erciyes University, Kayseri 38030, Turkey}
\newcommand{\FNAL}{Fermi National Accelerator Laboratory, Batavia, 
Illinois 60510, USA}
\newcommand{\FSU}{Florida State University, Tallahassee, Florida 32306, USA}
\newcommand{\UFG}{Instituto de F\'{i}sica, Universidade Federal de 
Goi\'{a}s, Goi\^{a}nia, Goi\'{a}s, 74690-900, Brazil}
\newcommand{\Guwahati}{Department of Physics, IIT Guwahati, Guwahati, 781 
039, India}

\newcommand{\Homi}{Homi Bhabha National Institute, Training School Complex,
 Anushakti Nagar, Mumbai 400094, India}
\newcommand{\Houston}{Department of Physics, 
University of Houston, Houston, Texas 77204, USA}
\newcommand{\IHyderabad}{Department of Physics, IIT Hyderabad, Hyderabad, 
502 205, India}
\newcommand{\Hyderabad}{School of Physics, University of Hyderabad, 
Hyderabad, 500 046, India}
\newcommand{\IIT}{Illinois Institute of Technology,
Chicago IL 60616, USA}
\newcommand{\Imperial}{Imperial College London, Department of Physics,
 London, United Kingdom}
\newcommand{\Indiana}{Indiana University, Bloomington, Indiana 47405, 
USA}
\newcommand{\INR}{Institute for Nuclear Research of Russia, Academy of 
Sciences 7a, 60th October Anniversary prospect, Moscow 117312, Russia}
\newcommand{\UIowa}{Department of Physics and Astronomy, University of Iowa, 
Iowa City, Iowa 52242, USA}
\newcommand{\ISU}{Department of Physics and Astronomy, Iowa State 
University, Ames, Iowa 50011, USA}
\newcommand{\Irvine}{Department of Physics and Astronomy, 
University of California at Irvine, Irvine, California 92697, USA}

\newcommand{\Magdalena}{Universidad del Magdalena, Carrera 32 No 22-08 Santa Marta, Colombia}
\newcommand{\MSU}{Department of Physics and Astronomy, Michigan State 
University, East Lansing, Michigan 48824, USA}

\newcommand{\Duluth}{Department of Physics and Astronomy, 
University of Minnesota Duluth, Duluth, Minnesota 55812, USA}
\newcommand{\Minnesota}{School of Physics and Astronomy, University of 
Minnesota Twin Cities, Minneapolis, Minnesota 55455, USA}
\newcommand{\Mississippi}{University of Mississippi, University, Mississippi 38677, USA}
\newcommand{\NISER}{National Institute of Science Education and Research, 
Bhubaneswar, Khurda, Odisha 752050, India}
\newcommand{\OSU}{Department of Physics, Ohio State University, Columbus,
Ohio 43210, USA}

\newcommand{\Panjab}{Department of Physics, Panjab University, 
Chandigarh, 160 014, India}
\newcommand{\Pitt}{Department of Physics, 
University of Pittsburgh, Pittsburgh, Pennsylvania 15260, USA}
\newcommand{\QMU}{Particle Physics Research Centre, 
Department of Physics and Astronomy,
Queen Mary University of London,
London E1 4NS, United Kingdom}

\newcommand{\SAlabama}{Department of Physics, University of 
South Alabama, Mobile, Alabama 36688, USA} 
\newcommand{\Carolina}{Department of Physics and Astronomy, University of 
South Carolina, Columbia, South Carolina 29208, USA}

\newcommand{\SMU}{Department of Physics, Southern Methodist University, 
Dallas, Texas 75275, USA}

\newcommand{\Sussex}{Department of Physics and Astronomy, University of 
Sussex, Falmer, Brighton BN1 9QH, United Kingdom}
\newcommand{\Syracuse}{Department of Physics, Syracuse University,
Syracuse NY 13210, USA}

\newcommand{\Texas}{Department of Physics, University of Texas at Austin, 
Austin, Texas 78712, USA}
\newcommand{\Tufts}{Department of Physics and Astronomy, Tufts University, Medford, 
Massachusetts 02155, USA}
\newcommand{\UCL}{Physics and Astronomy Department, University College 
London, 
Gower Street, London WC1E 6BT, United Kingdom}
\newcommand{\Virginia}{Department of Physics, University of Virginia, 
Charlottesville, Virginia 22904, USA}
\newcommand{\WSU}{Department of Mathematics, Statistics, and Physics,
 Wichita State University, 
Wichita, Kansas 67260, USA}
\newcommand{\WandM}{Department of Physics, William \& Mary, 
Williamsburg, Virginia 23187, USA}
\newcommand{\Wisconsin}{Department of Physics, University of 
Wisconsin-Madison, Madison, Wisconsin 53706, USA}

\affiliation{\ANL}
\affiliation{\Atlantico}
\affiliation{\Bandirma}
\affiliation{\BHU}
\affiliation{\Caltech}
\affiliation{\CUSB}
\affiliation{\Charles}
\affiliation{\Cincinnati}
\affiliation{\Cochin}
\affiliation{\CSU}
\affiliation{\CTU}
\affiliation{\Delhi}
\affiliation{\Erciyes}
\affiliation{\FNAL}
\affiliation{\FSU}
\affiliation{\UFG}
\affiliation{\Guwahati}
\affiliation{\Homi}
\affiliation{\Houston}
\affiliation{\Hyderabad}
\affiliation{\IHyderabad}
\affiliation{\IIT}
\affiliation{\Imperial}
\affiliation{\Indiana}
\affiliation{\ICS}
\affiliation{\INR}
\affiliation{\IOP}
\affiliation{\UIowa}
\affiliation{\ISU}
\affiliation{\Irvine}
\affiliation{\JINR}
\affiliation{\Magdalena}
\affiliation{\MSU}
\affiliation{\Duluth}
\affiliation{\Minnesota}
\affiliation{\Mississippi}
\affiliation{\NISER}
\affiliation{\OSU}
\affiliation{\Panjab}
\affiliation{\Pitt}
\affiliation{\QMU}
\affiliation{\SAlabama}
\affiliation{\Carolina}
\affiliation{\SMU}
\affiliation{\Sussex}
\affiliation{\Syracuse}
\affiliation{\Texas}
\affiliation{\Tufts}
\affiliation{\UCL}
\affiliation{\Virginia}
\affiliation{\WSU}
\affiliation{\WandM}
\affiliation{\Wisconsin}

\author{S.~Abubakar}
\affiliation{\Erciyes}

\author{M.~A.~Acero}
\affiliation{\Atlantico}

\author{B.~Acharya}
\affiliation{\Mississippi}

\author{P.~Adamson}
\affiliation{\FNAL}

\author{N.~Anfimov}
\affiliation{\JINR}

\author{A.~Antoshkin}
\affiliation{\JINR}

\author{E.~Arrieta-Diaz}
\affiliation{\Magdalena}

\author{L.~Asquith}
\affiliation{\Sussex}

\author{A.~Aurisano}
\affiliation{\Cincinnati}

\author{A.~Back}
\affiliation{\Indiana}
\affiliation{\ISU}

% ...

\makeatletter
\ExplSyntaxOn
\nova_flush_pending_author:
\ExplSyntaxOff
\newtoks\savedauthortoks
\newtoks\savedaffiltoks
\savedauthortoks=\expandafter{\the\auth@toks}
\savedaffiltoks=\expandafter{\the\affil@toks}
\newcommand{\printauthorlist}{%
  \section*{The NOvA Collaboration}
  \addcontentsline{toc}{section}{The NOvA Collaboration}
  {\raggedright \the\savedauthortoks\par}
  \medskip
  \begin{list}{}{%
    \setlength{\leftmargin}{0.28cm}%
    \setlength{\labelsep}{0pt}%
    \setlength{\itemsep}{\affiliationsSep}%
    \setlength{\topsep}{-\parskip}}
  \itshape\small
  \the\savedaffiltoks
  \end{list}}
\auth@toks={}
\affil@toks={}
\affilfalse
\makeatother

\emailAdd{mayly.sanchez@fsu.edu}
\emailAdd{andrew.sutton@duke.edu}
\emailAdd{ahimmel@fnal.gov}

\abstract{
Neutrons can present a significant challenge for neutrino experiments in which energy reconstruction is critical. With the ability to escape detection completely and with a weak correlation between their kinetic energy and any eventual energy deposition, it is difficult to fully account for neutrons produced in neutrino interactions. This in turn leads to significant model dependence when evaluating neutron-related systematic uncertainties. The NOvA experiment is a long-baseline neutrino oscillation experiment with a high-statistics sample of antineutrino data collected by its near detector. We report an excess relative to data of simulated neutron candidates with low energy depositions when using standard Geant4 physics lists. The simulation excess is traced to an overabundance of secondary photons produced from interactions of neutrons with kinetic energy greater than \SI{20}{\mega\eV}. Improved agreement with data is obtained by applying the data-driven neutron-on-carbon \menate model for neutrons between \SI{20}{\mega\eV} and ${\sim}$\SI{100}{\mega\eV}. With \menate, the residual oversimulation is more uniform across the calorimetric neutron energy spectrum, suggesting possible overproduction of primary neutrons by the GENIE neutrino interaction generator. These results motivate the adoption of \menate-supplemented Geant4 simulation as the nominal simulation in the production of future \nova simulation.} 


\arxivnumber{XXXx.XXXX} 

\begin{document}

\crefname{figure}{Figure}{Figures}

\maketitle
\flushbottom

\section{Introduction}
\label{sec:intro}

Neutrino oscillation measurements are a key probe of neutrino masses, mixing, and symmetries. Current and future accelerator-based experiments~\cite{NOvA:2025hbg, T2K:2025yoy, NOvAT2K:2025wet, DUNE2022} use observations of \numutonumuParen and \numutonueParen to measure the mass splitting $|\Delta m^2_{32}|$, its sign (a question known as the neutrino mass ordering), the mixing angles $\theta_{23}$ and $\theta_{13}$, and the phase $\delta_{\mathrm{CP}}$, which may indicate CP violation in the neutrino sector. Measurement of these parameters using neutrino oscillation requires accurate determination of the interacting neutrino flavor and energy. Comparisons of neutrino and antineutrino oscillation rates as a function of energy are especially important for measurements of CP violation and mass ordering.

Neutrons produced in neutrino and antineutrino interactions can complicate the measurement of event energies and thus the inferred oscillation parameters. Final-state neutrons can carry a significant portion of the final-state energy, and since neutrons are electrically neutral, this energy can go largely or entirely undetected. Neutrons are observed via the secondary particles produced when they interact in the detector material. Most of the available cross-section measurements relevant to the production of these secondary particles are applicable only to neutrons with kinetic energies below \SI{20}{\mega\eV} \cite{Blokhin:2016kje, Zhigang:2020kel, Brown:2018jhj, Plompen:2020yhs, Osamu:2023ouf, Koning:2019fks}, and there are few measurements for energies from a few tens to a few hundreds of MeV \cite{DelGuerra:1975stf, Cecil:1979}, which are relevant to \nova. Hence, the reconstructed energy of neutrino events with neutrons is incorrect by an amount that is difficult to model and is, on average, different for neutrino and antineutrino interactions because of differing neutron production rates. In fact, neutron-related uncertainties are among the largest in both neutrino oscillation and cross-section measurements~\cite{NOvA:2021nfi, MINERvA:2023avz, MicroBooNE:2022cls, DUNE:TDR}.

The impact of neutron-related uncertainties can be mitigated if experiments can identify energy depositions associated with neutrons and benchmark the simulation of neutron production and propagation in situ. Here, an algorithm is presented that tags neutron activity within the \nova Near Detector (ND) linked to the primary neutrino interaction. Two convolutional neural networks (CNN) trained to identify energy depositions from various secondary particles resulting from neutron interactions are also introduced. The neutron-tagging algorithm and neutron-secondary-particle-identifying CNNs are applied to antineutrino interactions in \nova to explore the detection of neutrons and compare data to simulation. Finally, the implementation of an alternative neutron interaction model, \menate~\cite{menateCreation}, is presented, and is shown to yield better agreement with data than \nova's default Geant4~\cite{Agostinelli:2003yb} configuration (v4.10.04 with the \texttt{QGSP\_BERT\_HP} physics list).

\section{The \nova experiment}
\label{sec:nova}

\nova is a long-baseline neutrino oscillation experiment consisting of two functionally identical detectors situated along the Neutrinos at the Main Injector (NuMI) beam produced at Fermilab in Batavia, Illinois \cite{NumI_2016}. The ND is located 100~m underground at Fermilab, approximately 1 km from the neutrino production target. The Far Detector (FD) is located on the surface near Ash River, Minnesota, \SI{810}{\kilo\meter} from the target. Both detectors are centered 14.6~mrad off axis from the NuMI beam and receive a narrowband neutrino or antineutrino flux peaked at \SI{1.8}{\giga\eV}. The beamline is equipped with magnetic focusing horns used to sign-select the neutrino parents, resulting in a 92\% (93\%) pure $\bar{\nu}_{\mu}$ ($\nu_{\mu}$) beam. Motivated by the increased neutron production in antineutrino interactions owing to charge conservation, this analysis uses the \nova ND $\bar{\nu}_{\mu}$ beam data and simulation. 

The \nova detectors are composed of extruded PVC cells, each filled with liquid scintillator and containing a loop of wavelength-shifting fiber that transports light to an avalanche photodiode. The cells are approximately 6.6 cm × 3.9 cm in cross section and are combined to form planes that are alternately oriented along the vertical and horizontal directions to achieve 3D reconstruction. The liquid scintillator is a blend of about 95\% mineral oil and 5\% pseudocumene and a small amount of a wavelength-shifting fluor~\cite{Mufson:2015kga}. The resulting detector composition by mass is 67\% carbon, 16\% chlorine, 11\% hydrogen, 3\% titanium, and 3\% oxygen. 

According to simulation, neutrons produced in the antineutrino interactions observed by \nova have kinetic energies peaked around \SI{15}{\mega\eV}, but with a long tail that pulls the mean kinetic energy up to about \SI{60}{\mega\eV}. These neutrons interact within the \nova detectors in three main ways: elastic scattering, inelastic scattering, and neutron capture. In elastic scattering, the recoiling nucleus produces scintillation light, while in inelastic scattering, secondary particles are created that then interact in the detector. These secondary particles include gamma emissions arising from nuclear de-excitations as well as charged hadrons such as protons or alpha particles.

Below about \SI{5}{\mega\eV}, neutrons are unable to scatter inelastically because of the threshold Q-values of the relevant processes. Just above this threshold, the inelastic scattering cross section rises rapidly to compete with the elastic process \cite{Brown:2018jhj}. If a neutron does not interact inelastically, elastic scatters eventually cause it to thermalize, leading to delayed neutron capture. In this analysis, reconstructed particle hits are required to be coincident in time with antineutrino events. Thus, simulation shows that the neutron interactions discussed here are split almost equally between elastic and inelastic scatters, with almost no observed neutron captures because of their intrinsic time delay.

\section{Simulation and neutron propagation modeling in NOvA}
\label{sec:models}

Neutrino interactions in \nova are simulated by GENIE v3.0.6 \cite{Andreopoulos:2009rq,Andreopoulos:2015wxa} with a custom model configuration that is then tuned to \nova ND data \cite{NOvA:2021nfi}. GENIE models quasielastic scattering (QE), resonant pion production (RES), deep inelastic scattering (DIS), and multinucleon interactions producing two ``holes'' (2p2h). Once these processes take place, GENIE further models final state interactions (FSI) where the outgoing particles can re-interact with the nuclear medium before emerging. The outgoing particles from GENIE are passed along to Geant4\footnote{Namely, v4.10.04 with a custom patch to more accurately calculate the density effect correction to the Bethe formula. This patch was directly incorporated in later Geant4 releases.}~\cite{Agostinelli:2003yb} and finally the energy depositions are input to a custom light model and electronics simulation to determine the detector response. 

Geant4 uses various interaction models for each particle over defined energy ranges. \nova uses the \texttt{QGSP\_BERT\_HP} physics list with no modifications to the default model configurations. The \texttt{QGSP\_BERT\_HP} physics list simulates incoming neutrons that interact inelastically via the following models:\footnote{At energies where models overlap, Geant4 mixes between them linearly.}
\begin{itemize}
     \item Neutron high precision (HP): \SIrange{0}{20}{\mega\eV}.
     \item Bertini intranuclear cascade (BERT): \SI{19.9}{\mega\eV}--\SI{9.9}{\giga\eV}.
    \item Fritiof (FTF): \SI{9.5}{\giga\eV}--\SI{25}{\giga\eV}.
    \item Quark Gluon String (QGS): \SI{12}{\giga\eV}--\SI{100}{\tera\eV}.

\end{itemize}
With a neutrino flux peaked at about \SI{2}{\giga\eV}, the most important models for neutrons in \nova are the Bertini cascade for energies above \SI{20}{\mega\eV} and the HP model below \SI{20}{\mega\eV}. 

The Bertini intranuclear cascade is further segmented into subprocesses within Geant4 \cite{G4PhysMan:2017aa}. First, the projectile interacts with a nucleon in the target nucleus. Then the secondary particles produced in that interaction are tracked through the nuclear medium, where subsequent interactions may occur or particles may escape the nucleus if their energies are high enough. These additional interactions cease once none of the remaining particles has enough energy to exit the nucleus. The residual nucleus is left in an excited state that is passed to Geant4's ``pre-compound''/``pre-equilibrium'' model and can emit nuclear fragments as the nuclear excitations are brought to equilibrium. Next, the excited nucleus is passed to the internal evaporation model associated with the Bertini cascade\footnote{The Fermi break-up model may also be used for nuclei lighter than $^{12}$C, but given the detector composition, these cases are rare in \nova.}. Six massive particles ($n$, $p$, $d$, $t$, $^3$He, and $\alpha$) are statistically emitted following Dostrowski's implementation of the model developed by Weisskopf and Ewing \cite{Dostrovsky:1959zz,PhysRev.118.791}. The photon-emission chain occurs only when no heavy particles can be produced. The built-in evaporation model for the Bertini cascade uses a constant density of energy levels for all nuclei and therefore does not reproduce all of the discrete level transitions of excited nuclei.

Low-energy neutrons (KE below \SI{20}{\mega\eV}) are simulated using the HP model. This model uses nuclear data tabulated in a neutron data library that combines measurements of interaction cross sections and final-state particle content \cite{Chadwick:2011endf}. These data are available only for some nuclei and typically only up to a few tens of MeV.

A supplementary model, \menate \cite{menateCreation}, has been investigated by the Modular Neutron Array (MoNA) collaboration for neutron-on-carbon inelastic scattering of neutrons produced from $^{16}\mathrm{B}$ decay \cite{ref:mona}. Like the built-in Geant4 HP model, \menate is data based and relies on cross-section measurements to provide a list of possible final-state particles. The available neutron-scattering datasets for most target isotopes cut off at an upper kinetic energy limit of about \SI{20}{\mega\eV}, thus this is the upper limit of the HP model. However, neutron-on-carbon measurements have been made up to ${\sim}$\SI{100}{\mega\eV} \cite{Cecil:1979, Guerra:1976}. Based on these measurements, \menate uses individual final-state cross sections and restricts the set of outgoing particle configurations while reproducing a total neutron-on-carbon inelastic cross section that is in good agreement with the one used by Geant4. The six possible channels produced by \menate are:
\begin{multicols}{2}
\begin{itemize}
    \item $^{12}\mathrm{C} + \mathrm{n} \rightarrow\, ^{12}\mathrm{C}^* + \mathrm{n}^{'} + \gamma$
    \item $^{12}\mathrm{C} + \mathrm{n} \rightarrow\, ^{9}\mathrm{Be} + \alpha$ 
    \item $^{12}\mathrm{C} + \mathrm{n} \rightarrow\, \mathrm{n}^{'} + 3\alpha$ 
    \item $^{12}\mathrm{C} + \mathrm{n} \rightarrow\, ^{12}\mathrm{B} + \mathrm{p}$
    \item $^{12}\mathrm{C} + \mathrm{n} \rightarrow\, ^{11}\mathrm{B} + \mathrm{p} + \mathrm{n}^{'}$ 
    \item $^{12}\mathrm{C} + \mathrm{n} \rightarrow\, ^{11}\mathrm{C} + 2\mathrm{n}^{'} $
\end{itemize}
\end{multicols}

\nova's implementation of \menate is inspired by the open-source NPTool simulation package \cite{NPTool:2016aa}, with some notable changes. In Geant4, each projectile has a list of associated processes. As the particle is transported through the medium, these processes compete with one another, and the one with the lowest interaction length is chosen for that step. The C++ implementation in NPTool \menate\footnote{Originally, \menate was implemented in FORTRAN. The Geant4-integrated C++ version was dubbed \menater; however, ``\menate'' is used here in all cases for simplicity.} was developed using a \textit{G4VDiscreteProcess}, which works well if all neutron simulations over the relevant energy range are handled exclusively by \menate. However, the \nova detectors are composed of additional elements, such as chlorine and titanium, which are not implemented in \menate. Thus, to preserve neutron interactions with these other nuclei, \nova cannot use a simple \textit{G4DiscreteProcess}.

Rather than implementing a new Geant4 process, \menate is introduced as an additional \textit{G4HadronicInteraction} model associated with the \textit{G4HadronicProcess} for neutrons. Doing so ensures that all desired neutron-on-carbon interactions are handled by \menate, while the built-in Geant4 intranuclear cascade models are used for all other targets and energies. Based on the cross sections used, the applicable energy range of the \menate model is restricted to \SIrange{20}{200}{\mega\eV}. For interactions on carbon, the Geant4 intranuclear cascade model is limited to kinetic energies above \SI{100}{\mega\eV}, so only \menate is used below that energy and down to the \SI{20}{\mega\eV} threshold. In the overlap region, Geant4 applies a linearly weighted superposition of the two models based on their respective cross sections.

\begin{figure}[t]
    \centering
    \subfloat[]{
        \includegraphics[width=.48\textwidth]{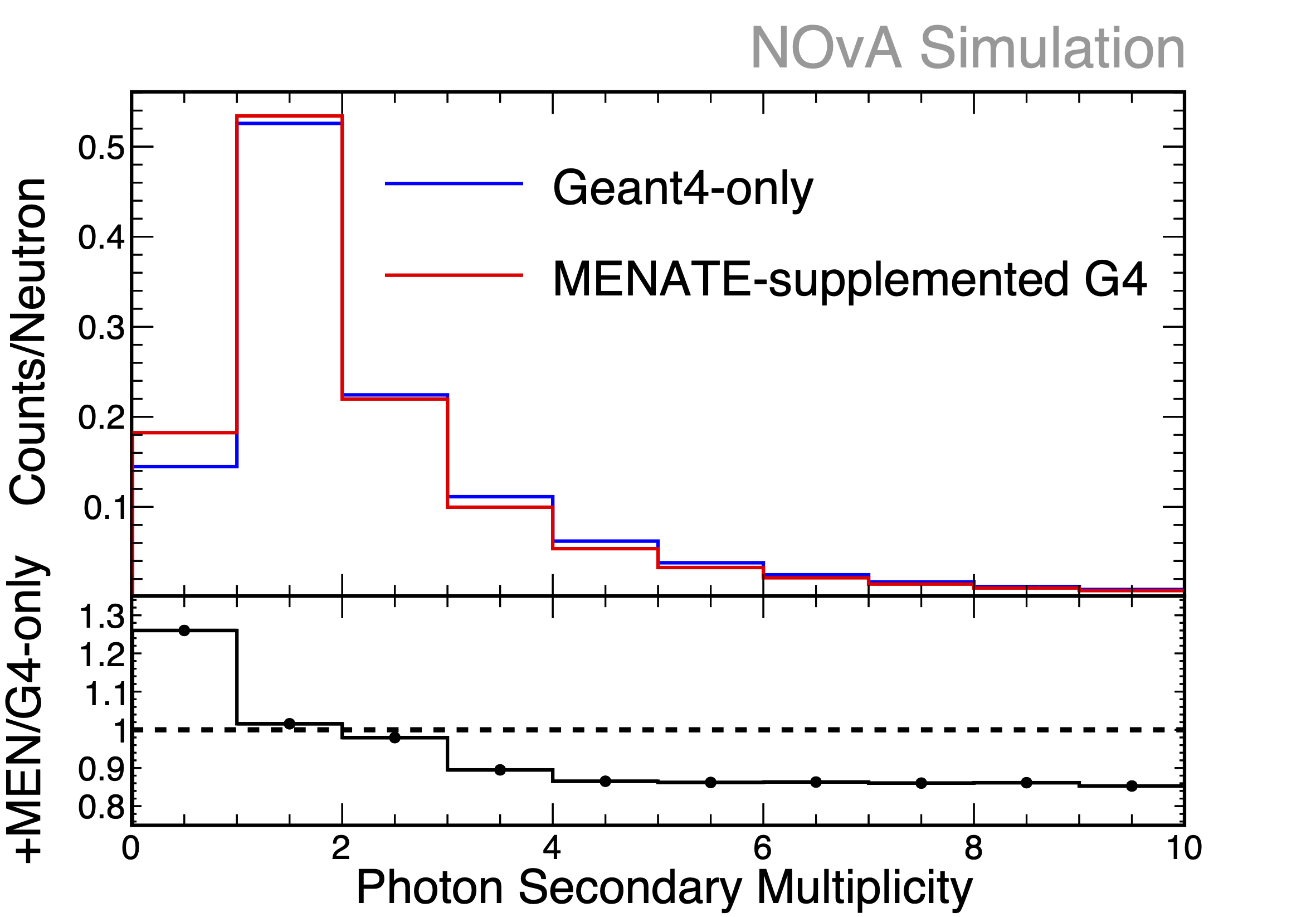}
        \label{fig:dau_mult_phot}}
    \subfloat[]{
        \includegraphics[width=.48\textwidth]{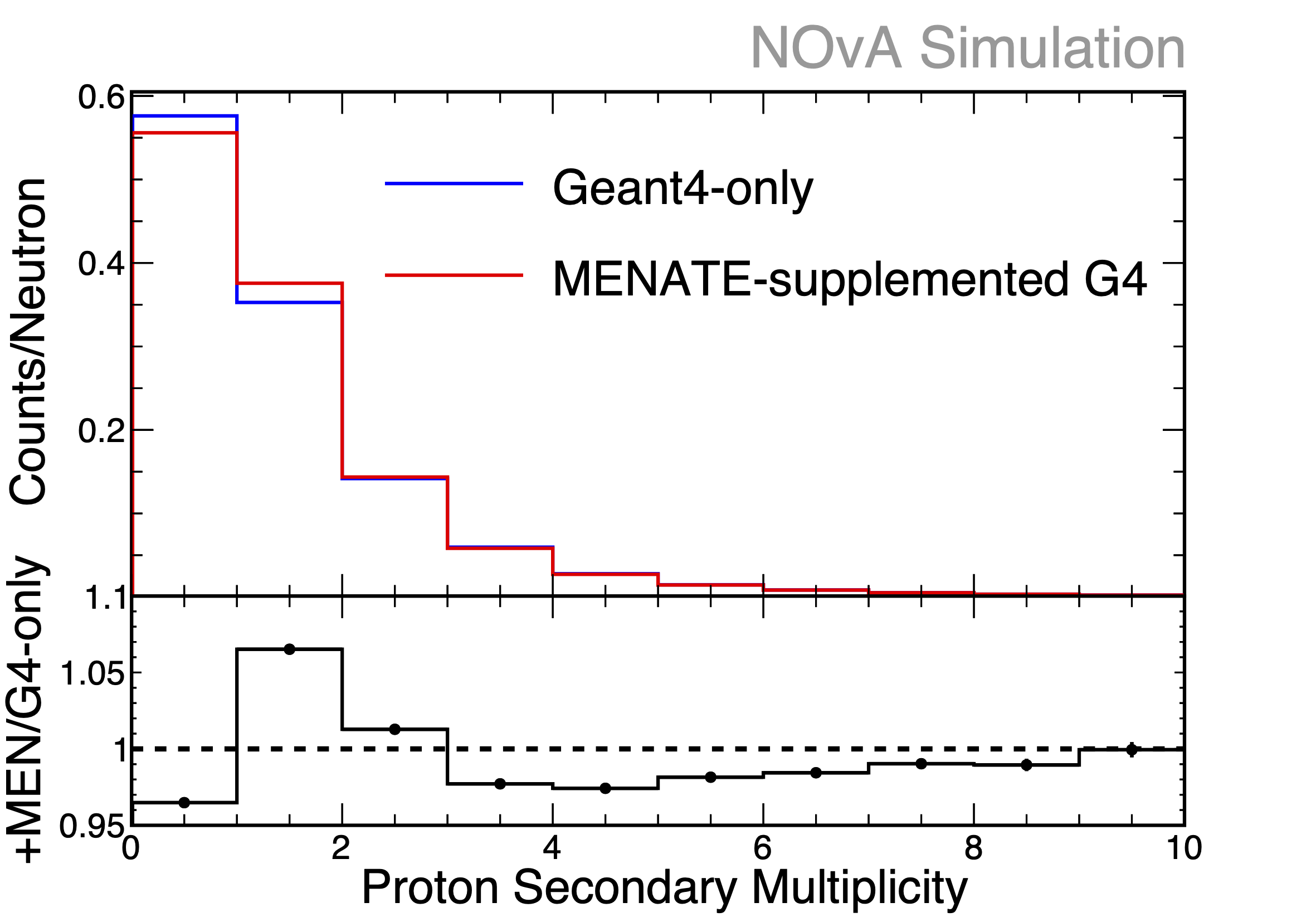}
        \label{fig:dau_mult_prot}}
    
    \caption{Simulated multiplicities of visible secondary (a) photons and (b) protons from primary neutrons that produced visible energy depositions. For Geant4-only (blue) and \menate-supplemented Geant4 (red). The statistical error bars are too small to be seen.}
    \label{fig:dau_mult}
\end{figure}

The result of using the cross section-driven model of \menate rather than the statistical intranuclear cascade of Geant4 is shown in Figure~\ref{fig:dau_mult}, which displays the simulated multiplicity of visible secondary photons and protons produced by a visible primary neutron. A particle is considered visible if any of its energy depositions produces light above threshold in a cell, while a neutron is considered visible if any of its secondary particles is visible.
With the inclusion of \menate, there is a significant reduction in the multiplicity of visible photon secondaries. Figure~\ref{fig:dau_mult_phot} shows about a 25\% increase in neutron interactions that produce no visible photons. Additionally, the \menate-supplemented simulation has a lower rate of multiphoton production per visible neutron primary than Geant4-only. Figure~\ref{fig:dau_mult_prot} shows that inclusion of \menate produces a slight increase in interactions that yield single protons and a corresponding reduction in both zero-proton and multiproton interactions.

\section{Neutron Candidate Identification}
\label{sec:selection}

\begin{figure}[ht]
    \centering
    \includegraphics[width=.7\textwidth]{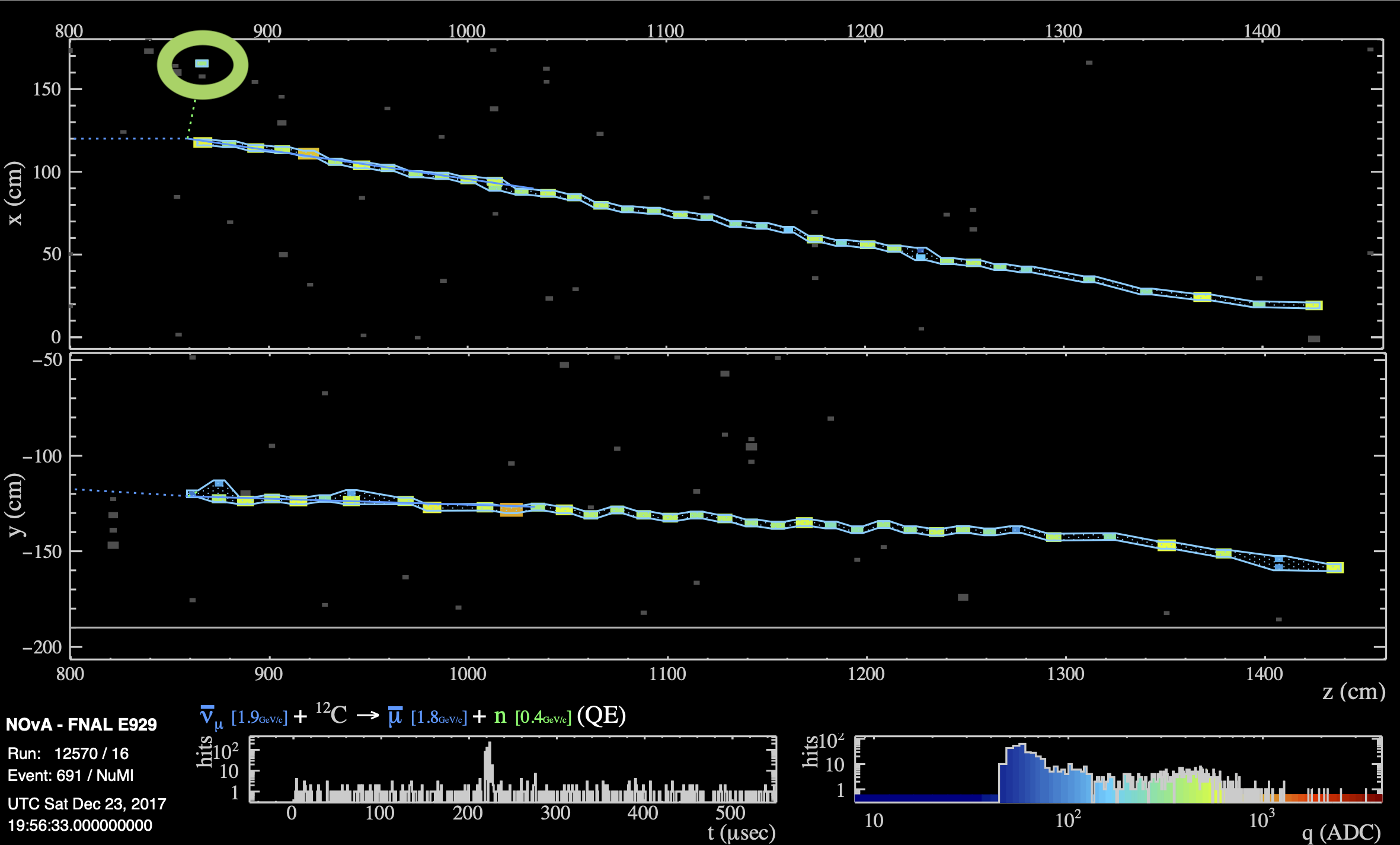}
    \caption{Simulated event showing a \SI{1.9}{\giga\eV} antineutrino interaction in the ND containing a \SI{400}{\mega\eV} primary neutron that leaves a small energy deposit (circled in green). The neutrino beam enters at left. The top half of the display shows hits registered in the $xz$ plane (i.e., viewed from above), while the bottom half shows hits in the $yz$ plane (i.e., viewed from the side).}
    \label{fig:event_display_simulation}
\end{figure}

Energy depositions above threshold are registered as detector ``hits.'' The threshold is determined on a cell-by-cell basis from the characteristic noise observed in the avalanche photodiode readout. The hits are grouped into ``prongs'' associated with single particles emanating from the reconstructed event vertex. 
The clustering procedure is performed separately for the orthogonal planes of the detector ($xz$ and $yz$ views), and the resulting two-dimensional prongs are matched between the two views to make three-dimensional prongs when possible. Prongs can take several forms, including a nearly straight track, a wider shower, or even just a few hits. Most prongs associated with primary neutrons are only two dimensional, with hits in only one of the two detector views, and are composed of only a few hits. Details of the NOvA reconstruction can be found in~\cite{NOvA:2021nfi}.

Of the simulated primary neutrons, only about 44\% go on to produce energy depositions that are reconstructed as a prong. The remaining neutrons are either completely unreconstructed (e.g., they exit the detector or their associated hits are below detection threshold) or their associated hits overlap, and are therefore indistinguishable from, hits of another particle. The selection criteria detailed below apply only to reconstructed prongs. This requirement introduces an effective threshold at about \SI{5}{\mega\eV} of neutron kinetic energy.

A simulated antineutrino event in the \nova ND is shown in Figure~\ref{fig:event_display_simulation}, which displays the $xz$ and $yz$ views of the detector with the neutrino beam entering from the left side. The long narrow track of hits visible in both views is a three-dimensional prong from the $\mu^{+}$ in the \numubar interaction. The small two-dimensional prong near the top of the $xz$ view (circled in green) is produced by a secondary particle from the neutron interaction.

\subsection{Identification criteria}
\label{ssec:selection}

To maximize the yield of neutrons for this analysis, antineutrino interactions are selected using criteria developed for the \nova three-flavor oscillation analysis, described in detail in Refs.~\cite{NOvA:2021nfi, NOvA:2019nfi}. The event sample consists of selected ND \numubar interactions. The longest prong in the event is most likely produced by a muon in the \numubar interaction and is therefore excluded. Additional criteria are applied to the other prongs in the event to isolate those associated with primary neutrons by taking advantage of the neutron's distinctive physical properties. These criteria are based on basic physics considerations and are largely independent of the particular interaction model employed.

\begin{figure}[htb]
    \centering
    \subfloat[]{\includegraphics[width=.49\textwidth]{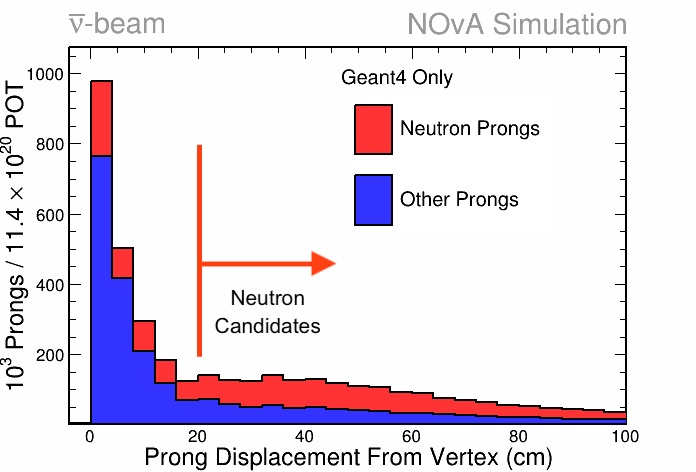}\label{fig:before_cuts_geant4_displ}}
    \subfloat[]{\includegraphics[width=.49\textwidth]{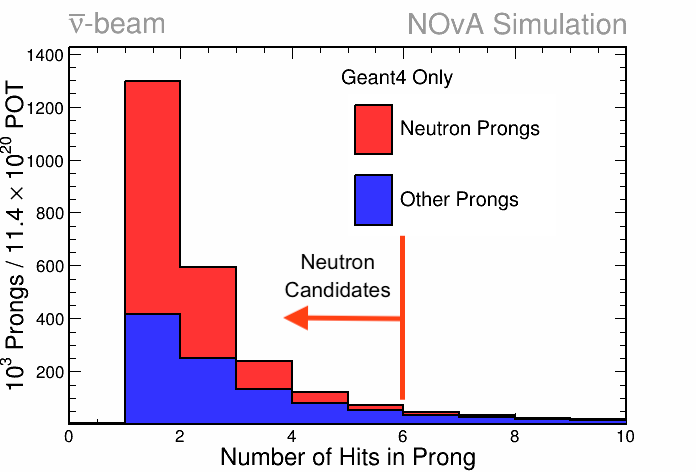}\label{fig:before_cuts_geant4_nhits}}
    \caption{ND antineutrino-mode particle prong distributions for simulations with the \numubar selection applied, with the muon prong candidate removed. (a) Displacement between the reconstructed neutrino interaction vertex and the starting point of the prong with the number of hits cut applied, (b) the number of hits in each prong with the displacement cut applied. The red histograms are prongs associated with a primary neutron; the blue histograms are prongs associated with non-neutron primaries. }
    \label{fig:before_cuts_geant4}
\end{figure}

In \nova, neutrons with kinetic energies up to \SI{150}{\mega\eV} have an average mean free path (MFP) of $\sim$\SI{30}{\centi\meter}. Such a large MFP means that prongs associated with neutrons are typically displaced farther from the neutrino interaction vertex than those associated with other particle types, as seen in Figure~\ref{fig:before_cuts_geant4_displ}. Thus, only prongs displaced more than \SI{20}{\centi\meter} from the neutrino interaction vertex are selected to greatly suppress non-neutron backgrounds. The remaining background contamination at large displacement is due mainly to $\pi^0$ production that yields secondary photons.

Unlike charged particles, neutrons are not directly detectable via ionization. Instead, they are visible only through the interactions of secondary particles produced in scattering or capture. The energies of secondary particles (e.g., protons, photons, and alpha particles) are kinematically limited, with the remaining nucleus taking away most of the available energy. Those nuclei are often below the detection threshold, so only the low-energy light secondary particles are visible. Because these secondary particles have low energies, they produce light in only a few cells. Thus, neutron candidates are required to have five or fewer hits, as seen in Figure~\ref{fig:before_cuts_geant4_nhits}.

\subsection{Neutron selection performance}
\label{ssec:performance}

\begin{figure}[ht]
    \centering
    \subfloat[]{\includegraphics[width=.60\textwidth]{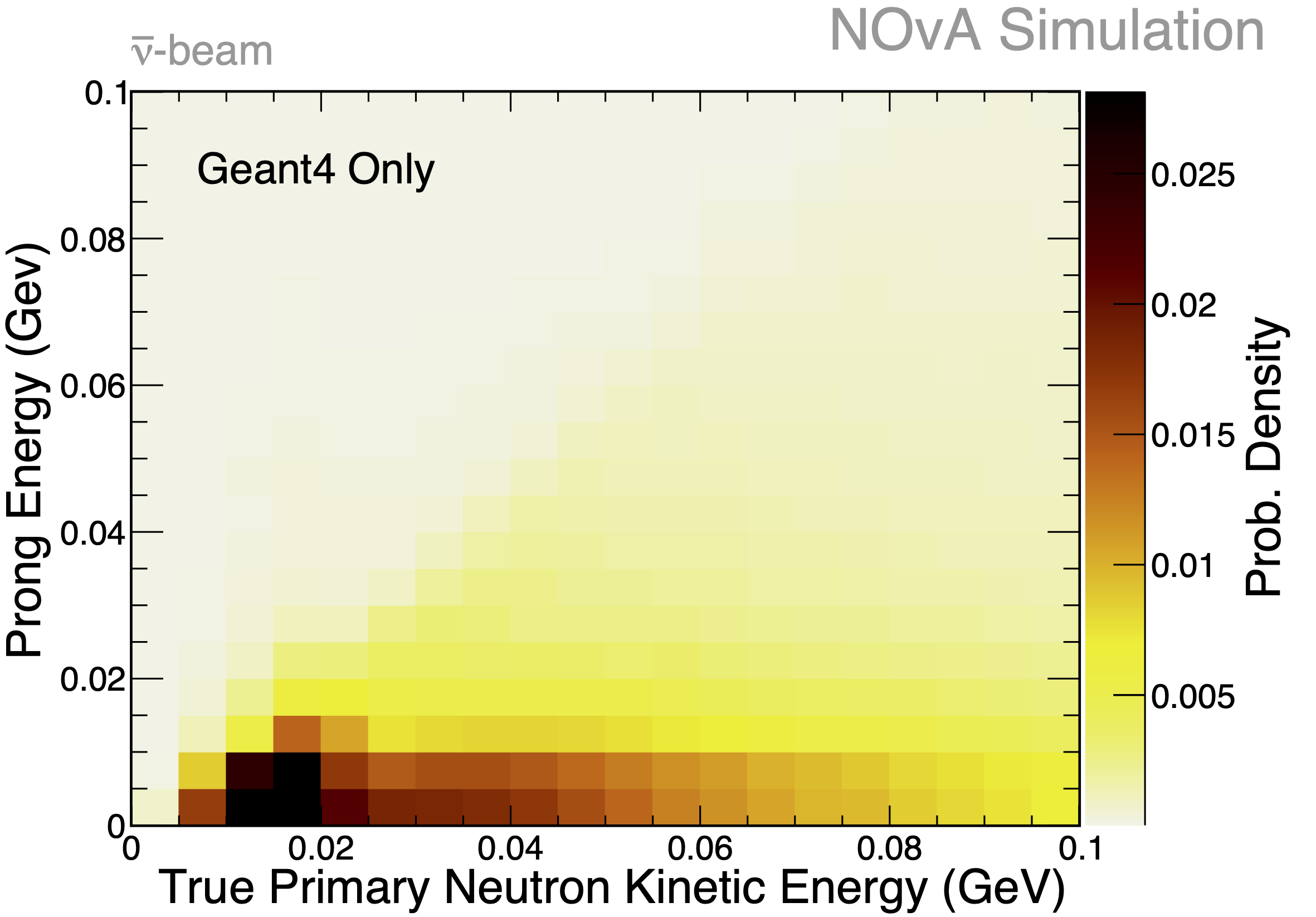}}    
    \caption{Probability density distribution of prong calorimetric energy vs the true kinetic energy of the prong's related primary neutron from \nova's default Geant4 simulation. If a primary neutron has more than one associated prong, their energies are summed.}
    \label{fig:2D_visE_trueE_geant4}
\end{figure}

The neutron candidate sample provides insight into the relationship between the energy deposited in the \nova ND and the true kinetic energy of the primary neutron. Figure~\ref{fig:2D_visE_trueE_geant4} shows the calorimetric energies of true neutron-associated prongs that pass the selection criteria versus the true kinetic energy of their primary neutron, where calorimetric energy is sum of the calibrated energies of the hits in the prong with a constant correction factor for the inactive material. When a single primary neutron is associated with more than one selected prong, the energies of those prongs are summed. The distribution demonstrates that there is little correlation between the kinetic energy of the primary neutron and the calorimetric energy deposited in the detector. In most cases, regardless of the neutron's kinetic energy, the visible energy deposited in prongs is less than about \SI{10}{\mega\eV}. This contrasts with the behavior of charged particles such as protons, pions, and muons, whose total energy deposition is well correlated with kinetic energy.

\begin{figure}[b]
\centering
    \subfloat[]{\includegraphics[width=.49\textwidth]{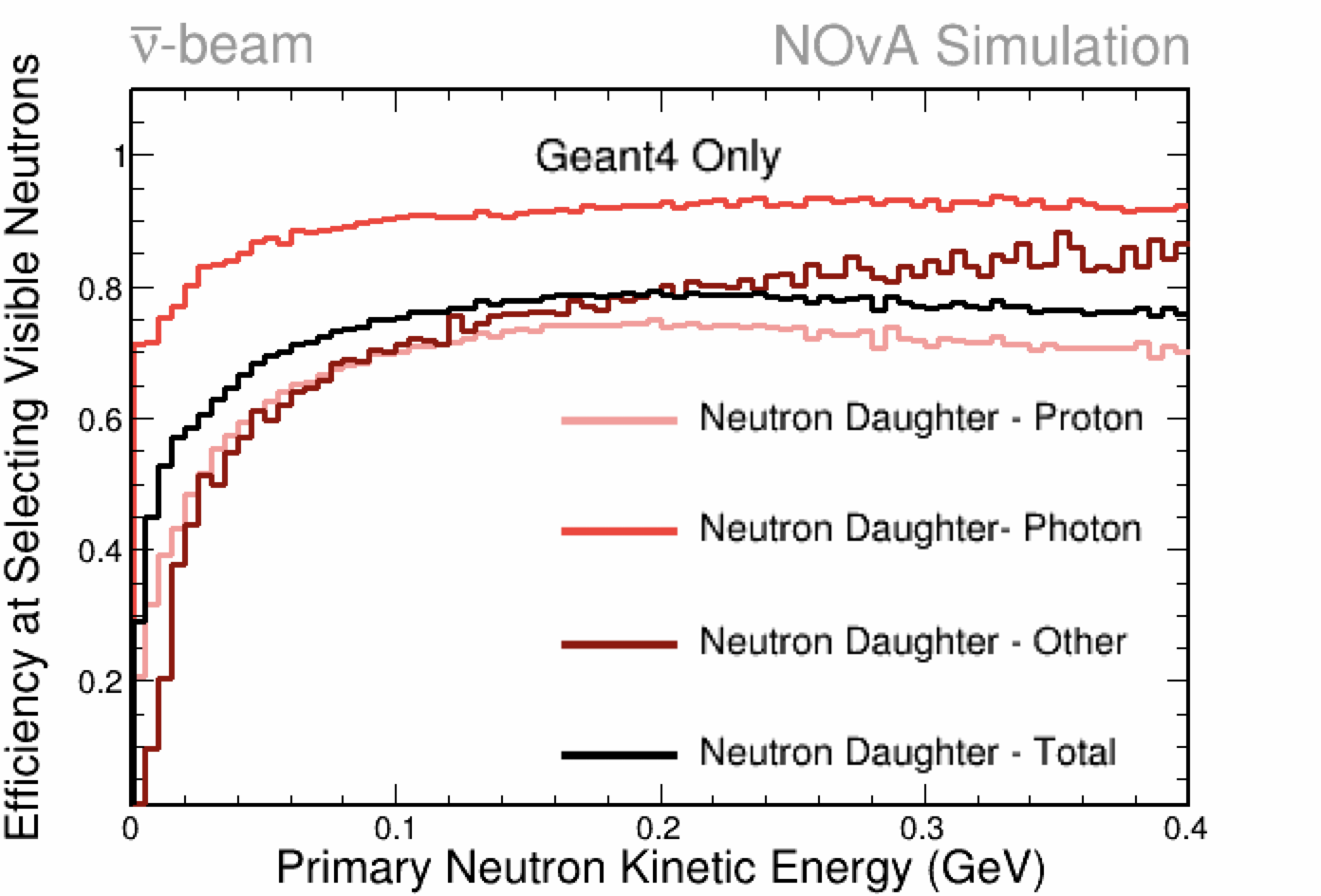}
    \label{fig:geant4_eff}}
    \subfloat[]{\includegraphics[width=.49\textwidth]{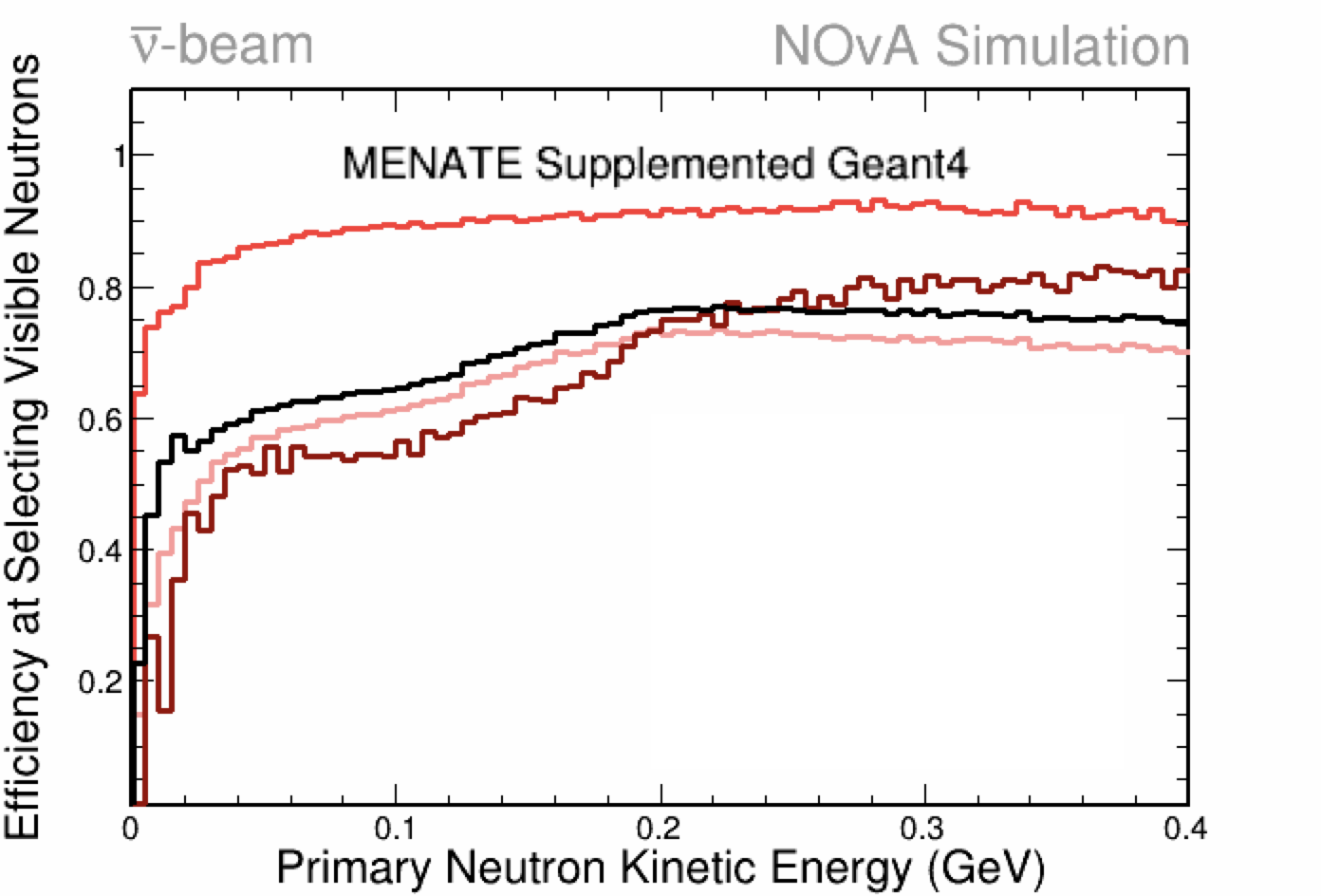}
    \label{fig:menate_eff}}
    \caption{Reconstructed neutron selection efficiency as a function of the kinetic energy of the primary neutron. The red histograms show the selection efficiencies split by the true secondary particle that produced the prong. (a) NOvA's default Geant4-only simulation, (b) the \menate-supplemented simulation. }
    \label{fig:eff}
\end{figure}

Figure~\ref{fig:geant4_eff} shows that while a prong's calorimetric energy is not well correlated to the parent neutron's kinetic energy, the ability to identify neutron-associated prongs is directly related to the kinetic energy of those neutrons. Higher energy neutrons are more likely to produce distinguishable prongs.
The selection efficiency also depends on the type of secondary particle that produces light, with secondary photons being the easiest to identify. Overall, the neutron prong selection is $71\%$ efficient and provides a sample of prongs that is $61\%$ pure in the nominal \nova $\bar{\nu}_{\mu}$ simulation using Geant4. The downturn of the selection efficiency at low energies in Figure~\ref{fig:geant4_eff} shows the effective \SI{5}{\mega\eV} detection threshold mentioned above.

Figure~\ref{fig:geant4_eff} shows an apparent plateauing at about 75\% selection efficiency for neutrons with high kinetic energy. However, new features arise when considering the visible calorimetric energy of the resulting prongs. About 10\% of all neutron-associated prongs, including those not selected by the neutron tagging algorithm, have \SI{100}{\mega\eV} or more of calorimetric energy. Although there is lack of good correlation between neutron kinetic energy and prong calorimetric energy, these prongs can only have been produced by the highest-energy neutrons in the sample. The selection efficiency for these high-energy prongs drops to about 50\%. One potential explanation is that there is a decrease in interaction cross section as the neutron's energy increases. The parent neutrons responsible for these high-energy secondaries will have a larger mean free path, and in turn a greater probability of escaping the detector unobserved.

The algorithm developed for neutron-prong selection was optimized using Geant4-only simulation. The selection remains robust when applied to the \menate-supplemented simulation; however, some degradation is seen in the selection efficiency. Comparing the plots in Figure~\ref{fig:eff} shows good agreement below \SI{20}{\mega\eV} and above \SI{200}{\mega\eV}, where Geant4-only is retained. Any discrepancies in these ranges reflect the statistical independence of the two simulation samples. At the upper limit of the \menate-only region (\SI{100}{\mega\eV}), the selection efficiency drops from 75\% in the Geant4-only sample to about 65\% when \menate is included. This drop is due to the reduced selection of non-photon secondaries. A re-optimization of the selection criteria could improve the performance on the \menate-supplemented simulation, though there is not much room for alterations in such a simple selection. Thus, it was decided to maintain selection consistency across the samples for the purpose of making direct comparisons.

\section{Comparison of simulations with \nova data}
\label{sec:results}

Neutrons can carry a significant amount of energy away from an antineutrino interaction; thus, significant biases are introduced when calorimetric energy is used to estimate the incoming antineutrino energy. The neutron-prong sample within the selected \nova ND $\bar{\nu}_{\mu}$ data provides insight into how well the \nova simulations describe visible neutrons, enabling an estimate of the size of the related uncertainty.

This sample also provides an opportunity to compare neutron-propagation models and assess which provides better agreement with the data. Antineutrino beam data collected between June 29, 2016, and February 26, 2019, are used here, corresponding to a total exposure of $11.4\times10^{20}$ protons on target \cite{NOvA:2019nfi}.

\subsection{Excess of neutron-induced photon secondaries produced by Geant4}
\label{ssec:g4datamc}

\begin{figure}[ht]
    \centering
    \subfloat[]{\includegraphics[width=.49\textwidth]{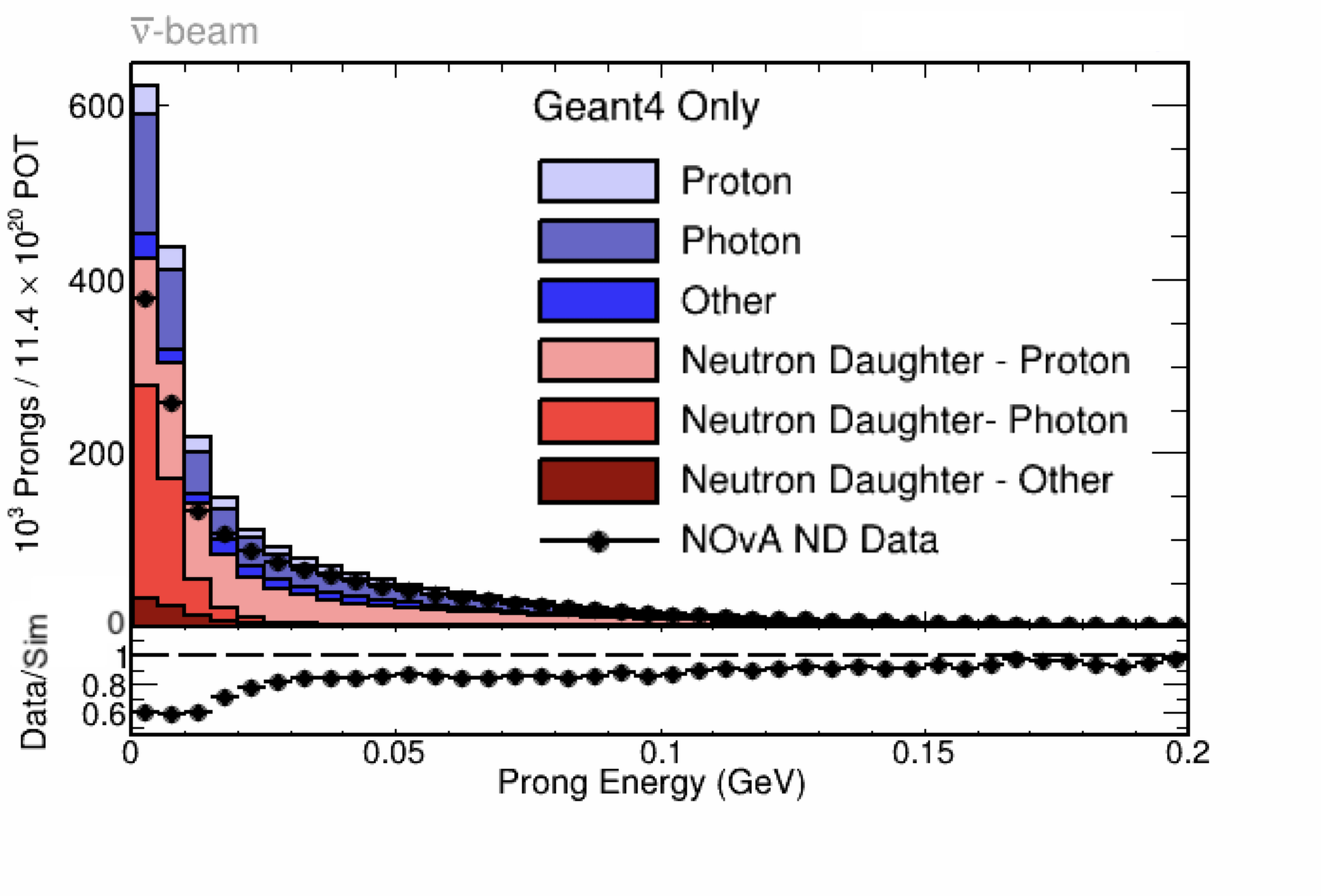}\label{fig:after_cuts_geant_a}}
    \subfloat[]{\includegraphics[width=.49\textwidth]{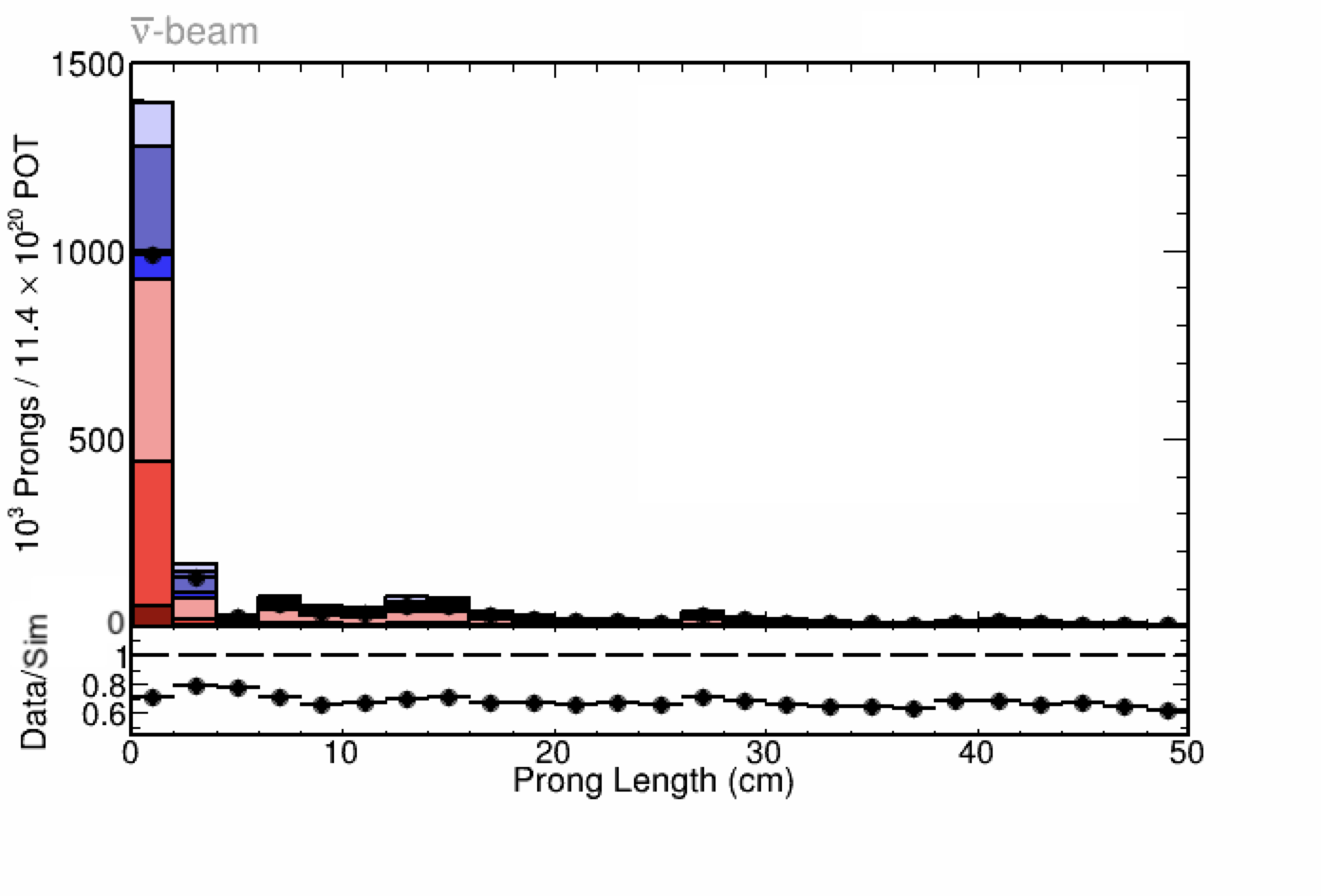}\label{fig:after_cuts_geant_b}}
    \\
    \subfloat[]{\includegraphics[width=.49\textwidth]{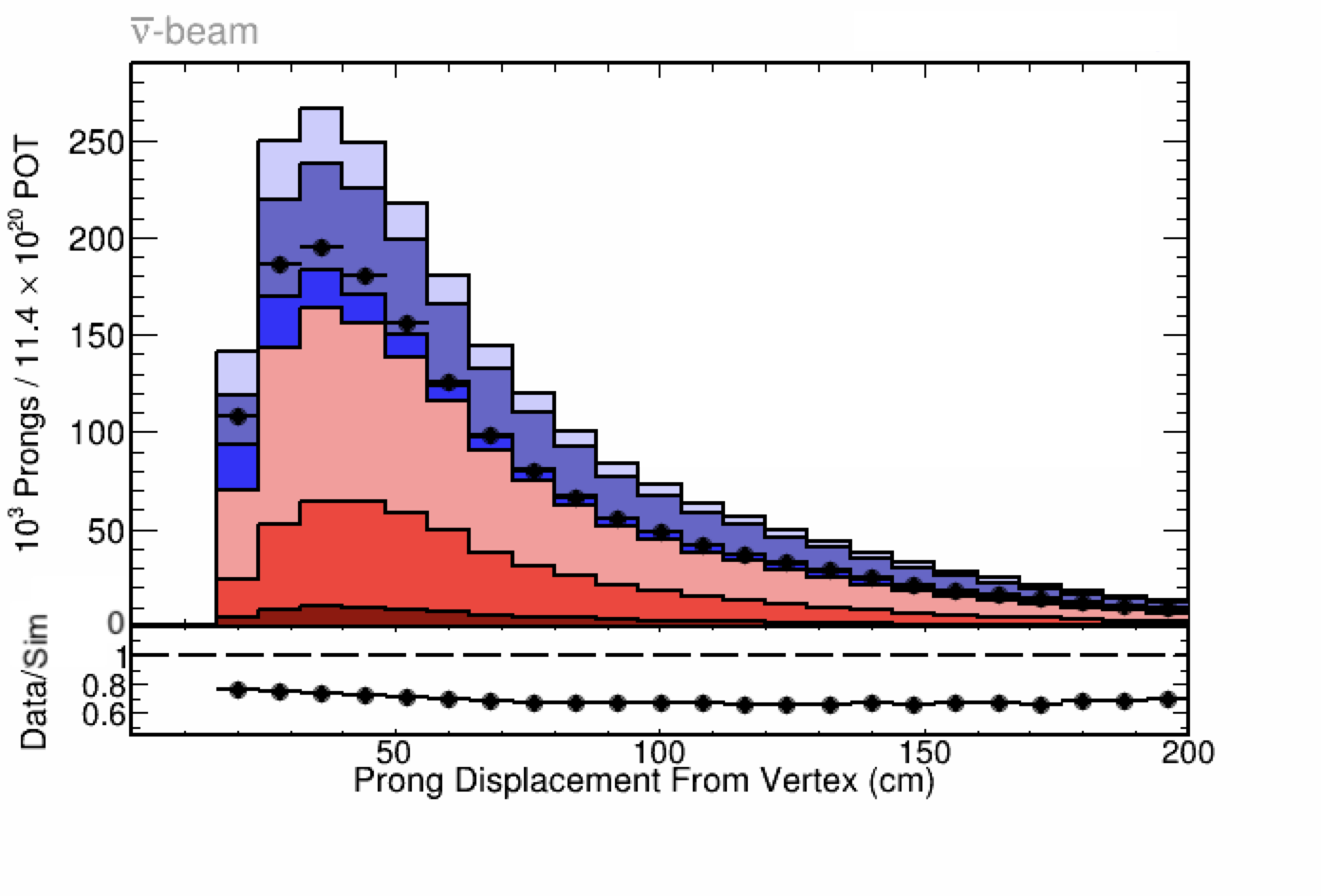}\label{fig:after_cuts_geant_c}}
    \subfloat[]{\includegraphics[width=.49\textwidth]{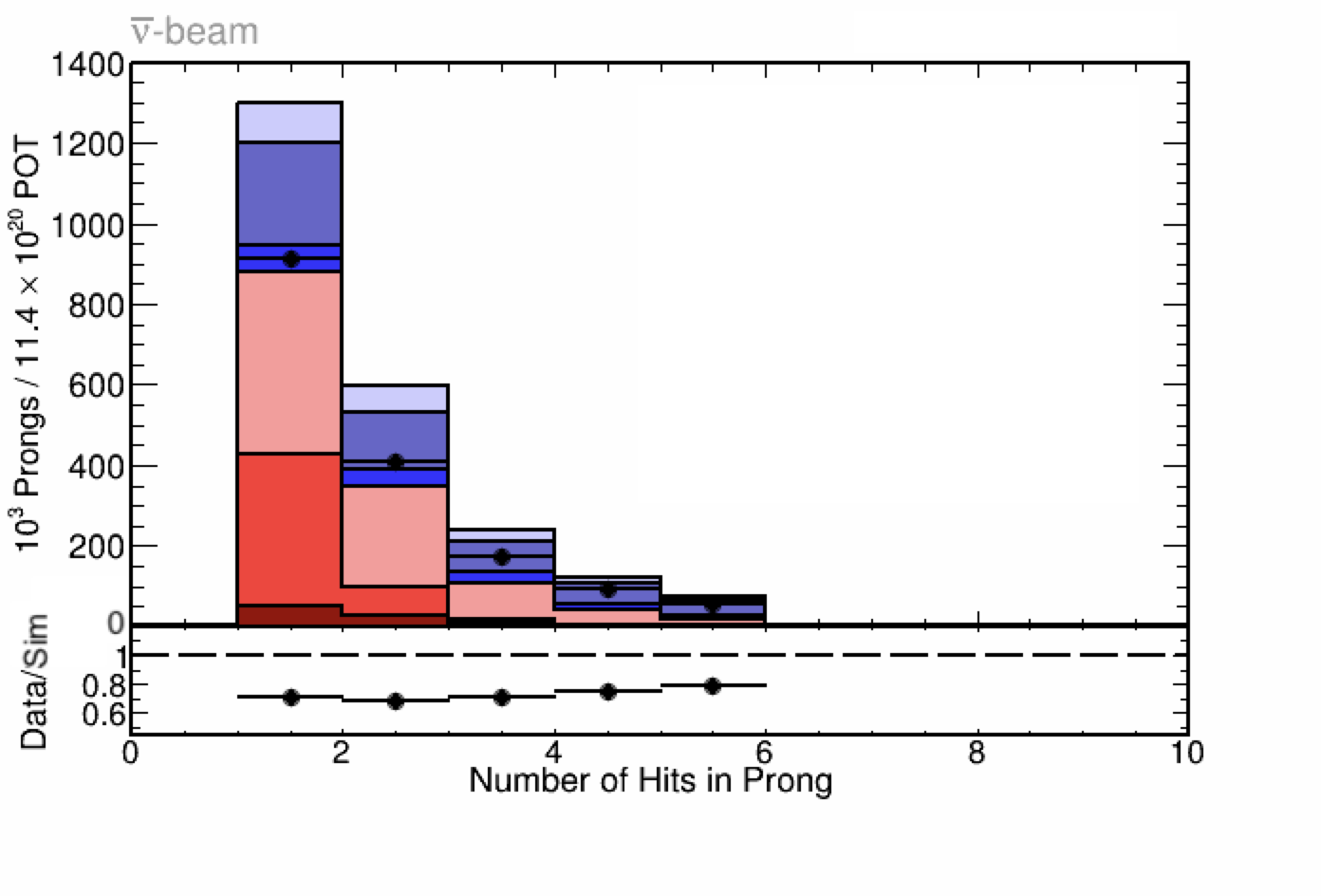}\label{fig:after_cuts_geant_d}}

    \caption{Distributions of neutron candidate prongs selected in \nova's default Geant4 simulation and ND antineutrino mode data along with data--simulation ratios for each distribution. (a) Calorimetric energy distribution, (b) reconstructed prong length, (c) displacement between reconstructed neutrino interaction vertex and prong start position, (d) number of hits in each selected prong. Red histograms are prongs associated with a primary neutron; blue histograms are prongs associated with non-neutron primaries.}
\label{fig:after_cuts_geant}
\end{figure}

The reconstructed energy distribution of the prongs selected by the neutron-tagging algorithm, Figure~\ref{fig:after_cuts_geant_a}, shows a low-energy excess of simulated prongs relative to the \nova ND $\bar{\nu}_{\mu}$ data. The simulation excess is most prevalent at the lowest prong energies, where about 40\% more selected prongs are observed below \SI{15}{\mega\eV}. The other reconstructed variables shown in \crefrange{fig:after_cuts_geant_b}{fig:after_cuts_geant_d} also display a simulation excess of \SIrange{20}{40}{\%}. The simulation shown in these plots is broken down by the true immediate particle that produced the prong, with the red histograms denoting prongs produced by secondary particles from neutron interactions and the blue histograms denoting contamination from particles that did not originate in a neutron interaction.

The prongs produced by secondary photons from a primary neutron interaction (labeled ``Neutron Daughter-Photon'' in the figure) appear mainly in the regions of the prong-energy distribution with the largest oversimulation. This suggests that the excess is at least partially related to the simulation of photon production from neutron interactions in the \nova ND. Of these neutron-produced photon prongs, about 88\% were generated from a parent neutron that was handled by Geant4's Bertini intranuclear cascade and produced via photon evaporation, implicating that particular model as a potential source of the observed oversimulation.

Additional insight is gained by using a convolutional neural network (CNN) trained to identify the types of secondary-particles that produce neutron-related prongs. A neutron CNN was trained on a sample of prongs produced by secondary particles from interactions of primary neutrons in \nova's Geant4-only simulation. The training sample used truth information to identify the prongs regardless of whether they would be selected by the neutron-tagging algorithm. As discussed above, neutron secondaries typically have low energies, resulting in hits in only a few cells. This means that there are often not enough hits to produce full three-dimensional reconstruction. Therefore, separate versions of the CNN were trained for two-dimensional and three-dimensional prongs because they have different topological characteristics. In the selected neutron-prong sample, 90\% of the prongs are two-dimensional.

The architecture used for the two neutron neural networks was developed from MobileNet Version 2 \cite{sandler2019mobilenetv2} and is the same architecture used in \nova to train other CNNs \cite{Psihas:2019ksa}. Here, the setup is modified slightly so that, in addition to the full event context, the isolated prong is also input, with both pixel maps split by detector view for four total inputs per prong. The neutron CNNs output a set of three scores, each ranging from 0 to 1, characterizing the probability that the prong resulted from a given particle type: photon, proton, or other secondary. The neutron-prong selection is dominated by secondary photons and protons, with low statistics for any ``other'' secondaries (heavier particles such as $d$, $\alpha$, and nuclear remnants). As a result, the CNNs have little power to identify the ``other'' secondaries and instead behave like binary classifiers between protons and photons.

\begin{figure}[t]
\centering
\subfloat[]{\includegraphics[width=.50\textwidth]
{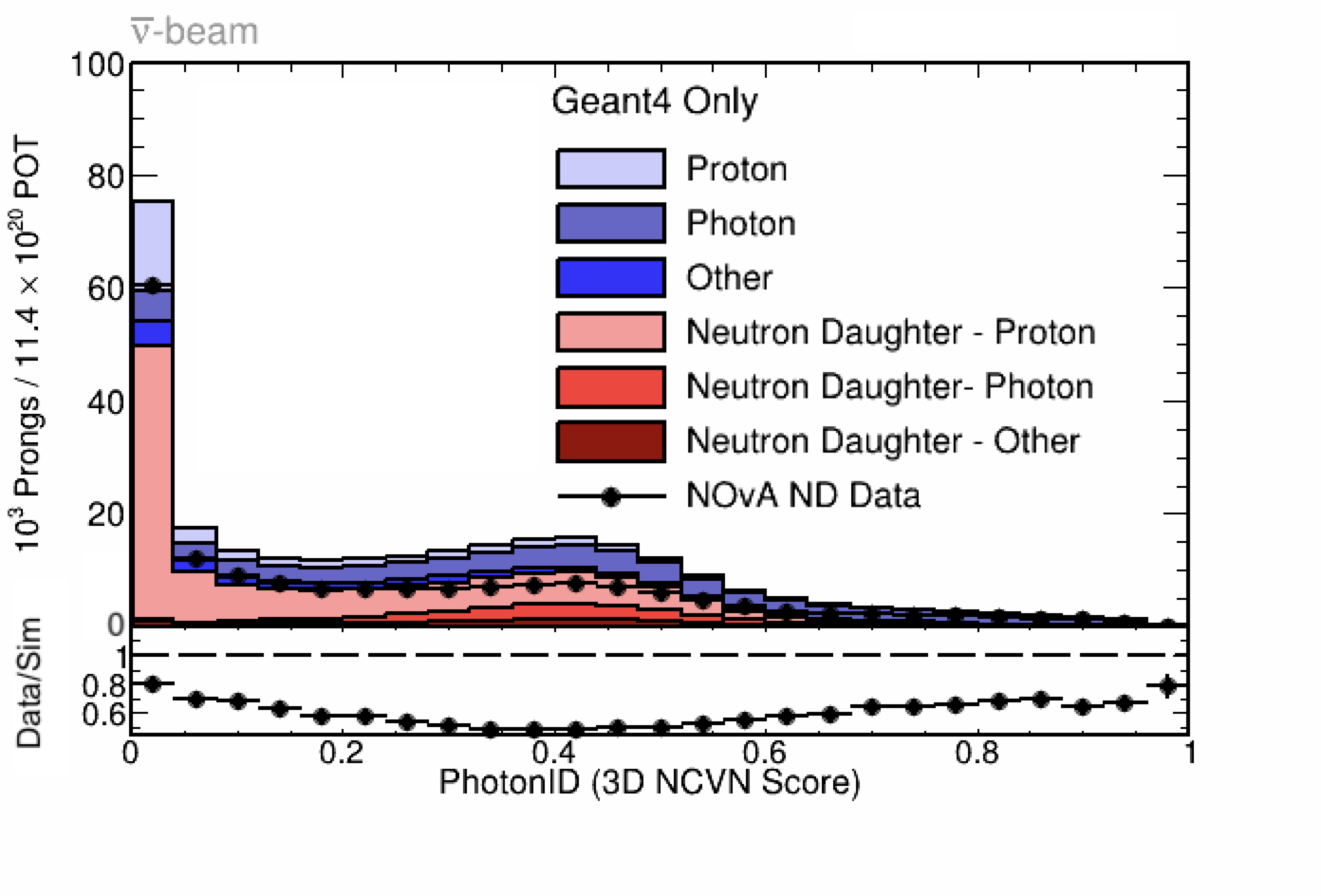}\label{fig:3dcvn_scores_geant}}
\subfloat[]{\includegraphics[width=.50\textwidth]
{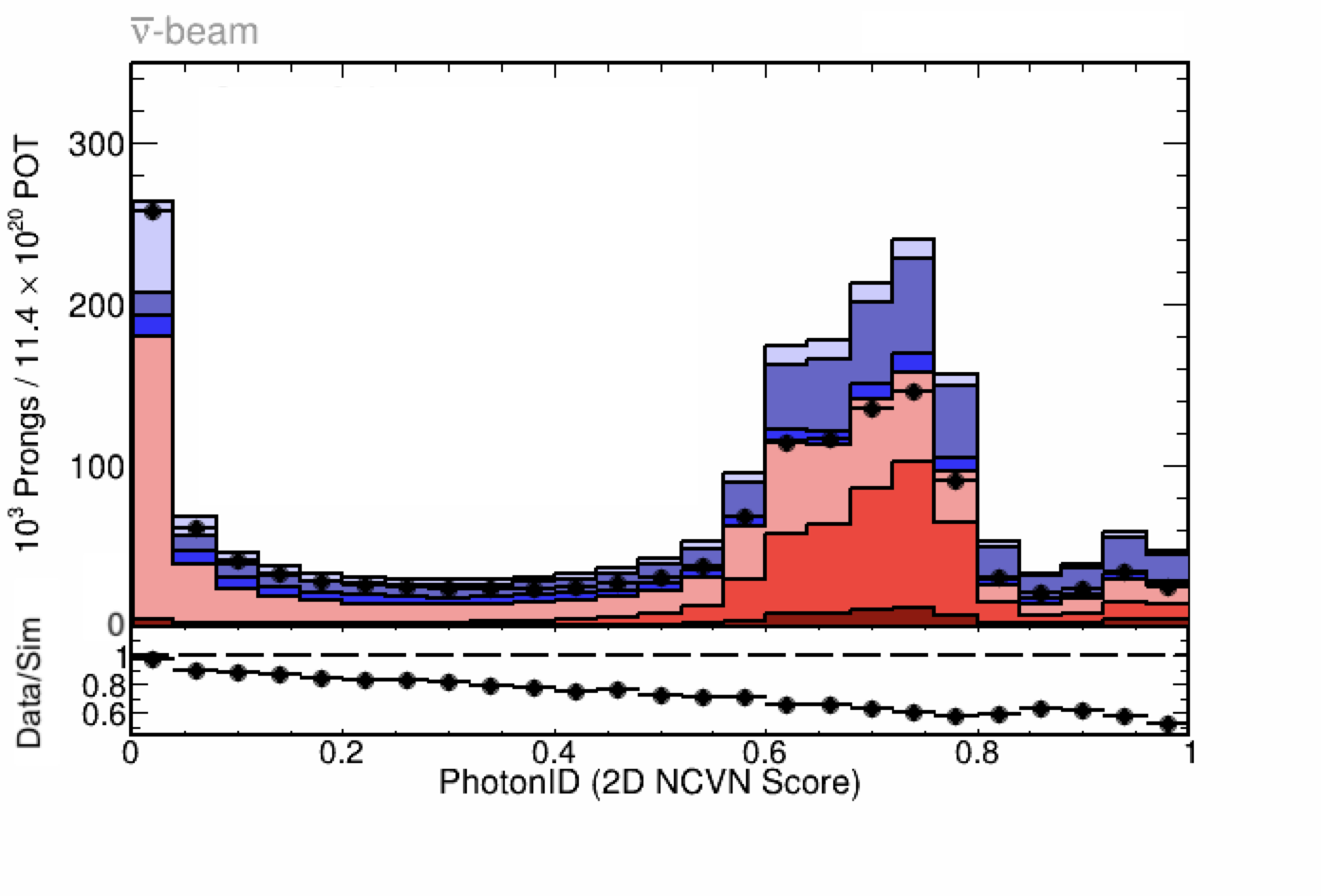}\label{fig:2dcvn_scores_geant}}
\caption{Photon identification scores from the neutron CNNs applied to neutron candidate prongs in data and \nova's default Geant4 simulation. (a) Three-dimensional prongs and (b) two-dimensional prongs.}
\label{fig:cvn_scores_geant}
\end{figure}

The neutron CNNs are used to investigate the prongs in \nova's default Geant4 sample selected by the neutron-tagging algorithm. The results of the two-dimensional-prong version with \nova's default Geant4 are shown in Figure~\ref{fig:2dcvn_scores_geant}. The peak at a photon score of zero, which corresponds to a proton score of nearly one, indicates that the two-dimensional CNN identifies proton-produced prongs well. For photon prongs, the CNN shows a wide peak at photon scores between 0.6 and 0.8 and a second, significantly smaller, peak at a photon score of 1. This separation of the signal region is likely due to photons resulting from differing interaction processes within the simulation. The majority of the selected three-dimensional prongs, shown in Figure~\ref{fig:3dcvn_scores_geant}, are produced by protons. Photons and other secondaries typically have energies that are too low to span the multiple detector planes required for three-dimensional reconstruction.

The best agreement with the data is seen in the proton-dominated bins of Figure~\ref{fig:cvn_scores_geant}, especially in the first bins. In contrast, the excess of simulated neutron candidates worsens as the secondary-photon content increases. In both the the two-dimensional and three-dimensional cases, the CNN distribution shows the largest data--simulation disagreement in the photon-populated regions. This provides additional evidence that one source of the excess lies in the photon-production component of the Geant4 model.

\subsection{\menate model improves agreement with the data}
\label{ssec:menatedatamc}

\begin{figure}[ht]
    \centering
    \subfloat[]{\includegraphics[width=.49\textwidth]{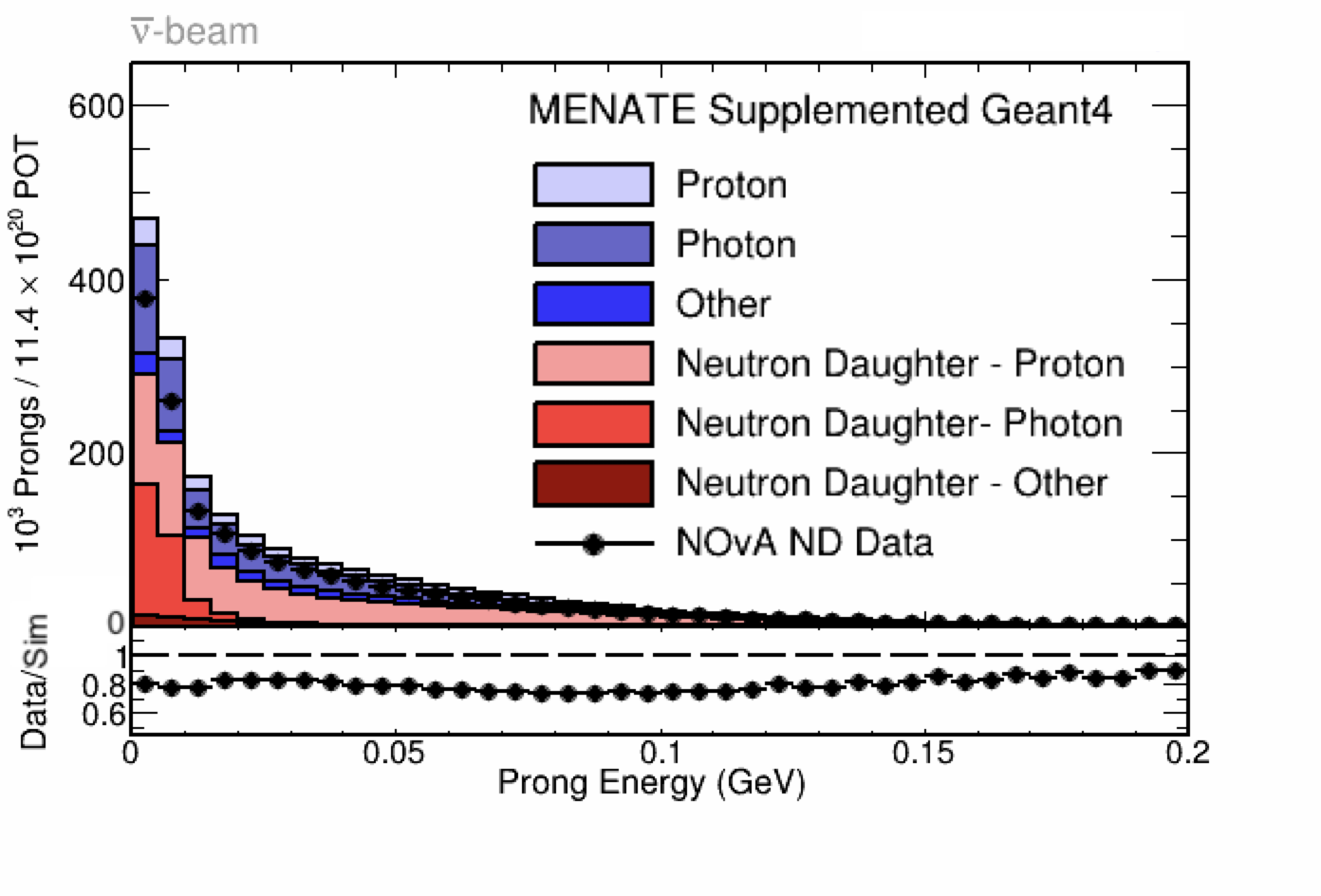}\label{fig:after_cuts_menate_a}}
    \subfloat[]{\includegraphics[width=.49\textwidth]{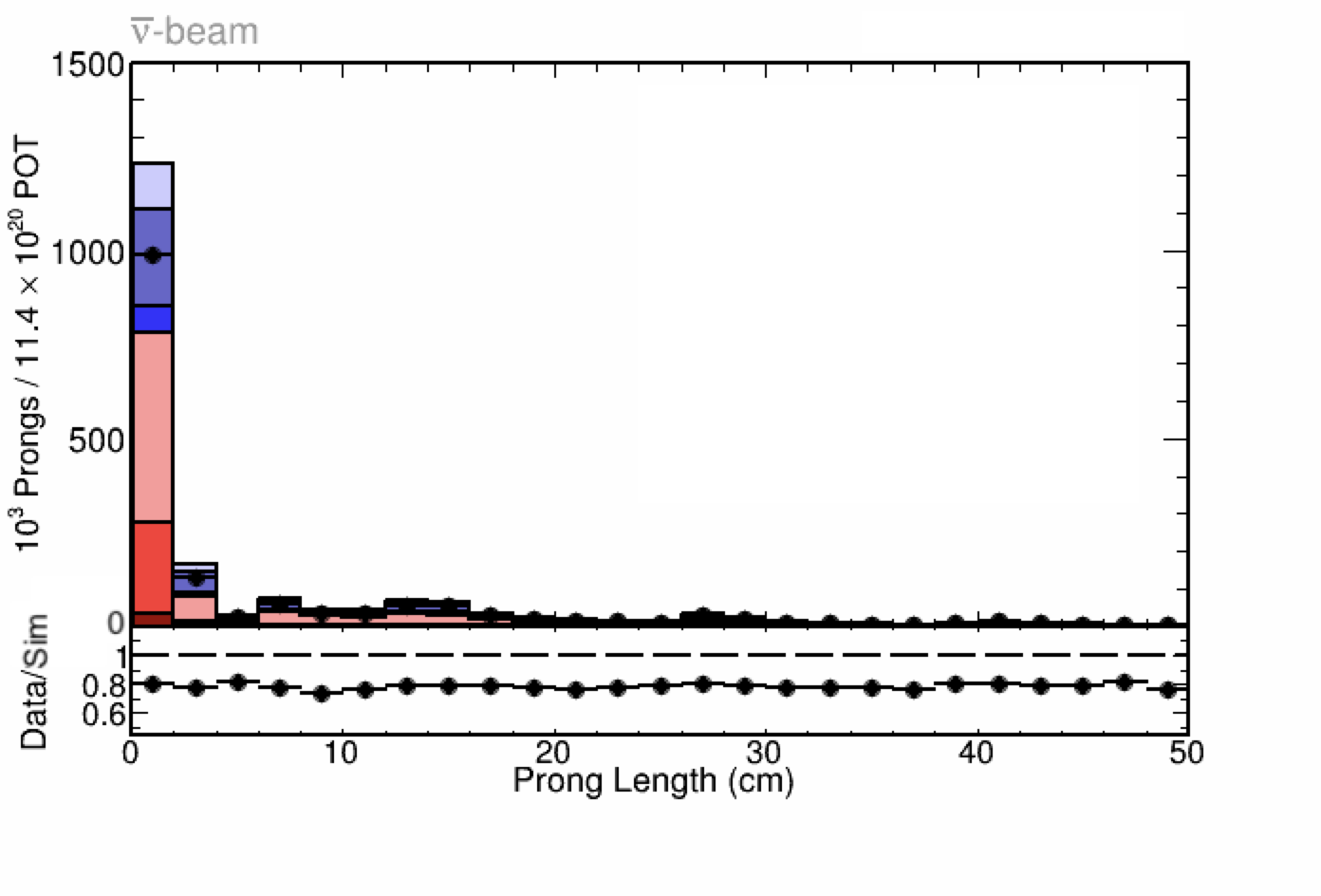}\label{fig:after_cuts_menate_b}}
    \\
    \subfloat[]{\includegraphics[width=.49\textwidth]{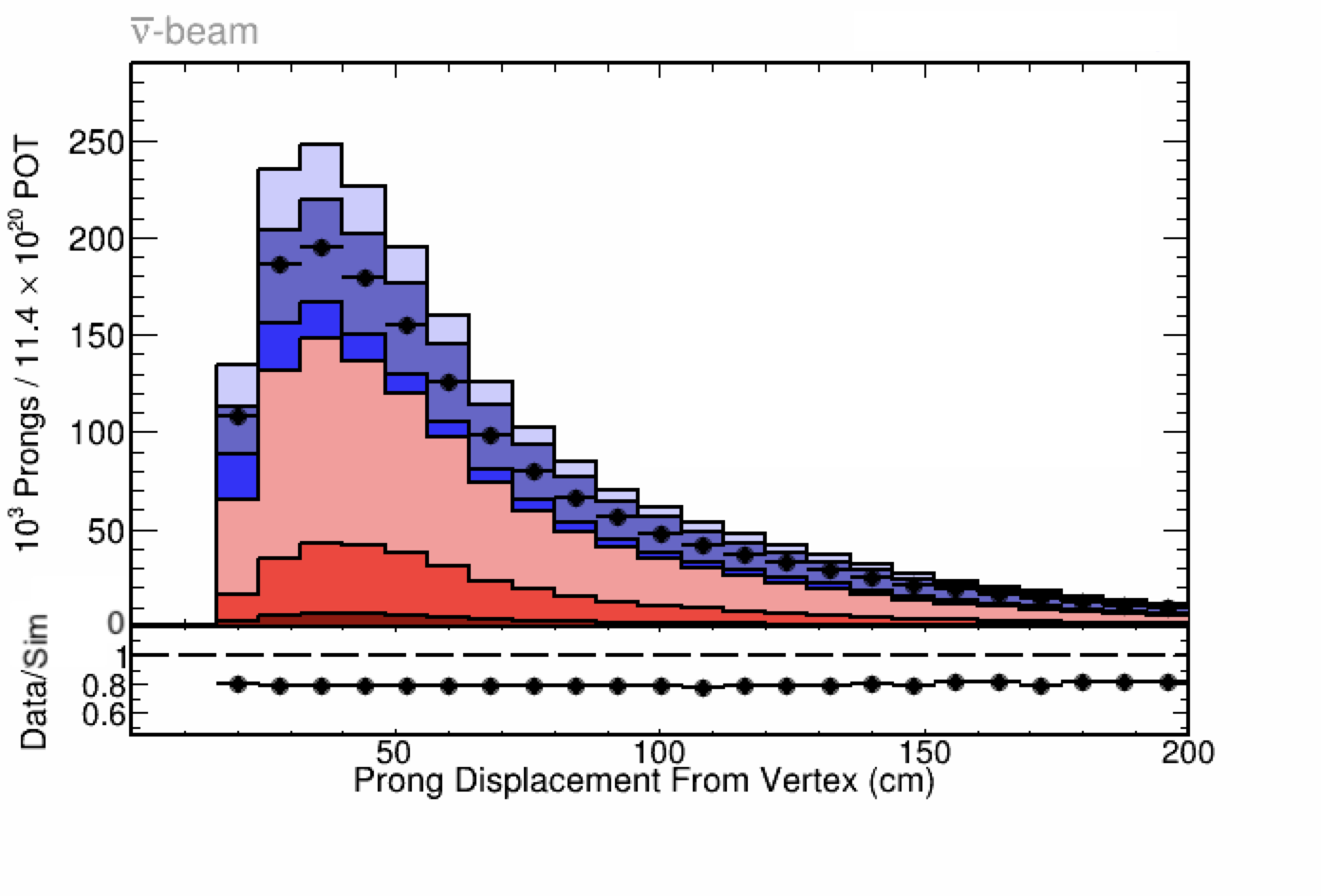}\label{fig:after_cuts_menate_c}}
    \subfloat[]{\includegraphics[width=.49\textwidth]{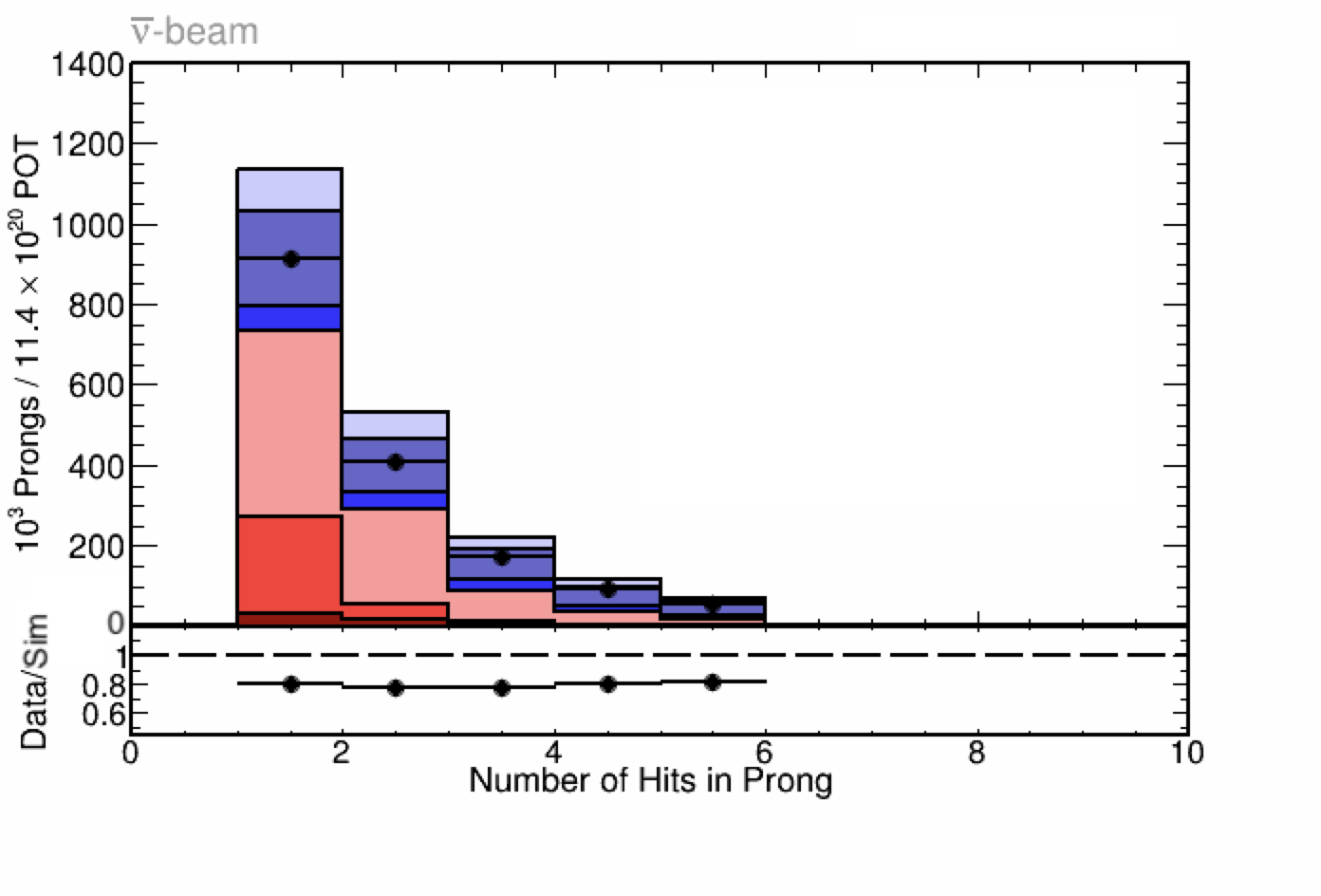}\label{fig:after_cuts_menate_d}}

    \caption{Distributions of neutron candidate prongs selected in the \menate-supplemented simulation and ND antineutrino mode data along with data--simulation ratios for each distribution. (a) Calorimetric energy distribution, (b) reconstructed prong length, (c) displacement between reconstructed neutrino interaction vertex and prong start position, (d) number of hits in each selected prong. Red histograms are prongs associated with a primary neutron; blue histograms are prongs associated with non-neutron primaries.}
\label{fig:after_cuts_menate}
\end{figure}

Simulated neutron-candidate prong distributions agree better with the \nova ND data when \menate is added to the Geant4 neutron-propagation model. The number of neutron-associated prongs produced by secondary photons decreases significantly, particularly at calorimetric energies below \SI{50}{\mega\eV}, as seen in Figure~\ref{fig:after_cuts_menate_a}. 
In addition to the decrease in neutron-induced photon prongs, the \menate-supplemented simulation shows a slight increase in secondary-proton prongs, which can be seen for prongs with energies greater than about \SI{15}{\mega\eV}. Figure~\ref{fig:dau_mult} displays the origin of both changes in secondary-prong content. \menate significantly reduces the multiplicity of photon secondaries, while at the same time shifting the distribution of secondary protons. The mean proton multiplicity is nearly unchanged between the models; however, \menate produces both fewer zero-proton and fewer multi-proton final states, instead favoring single-proton outcomes. When only a single proton is produced, it carries a larger fraction of the incoming neutron's energy, causing in turn a larger deposited energy and the shift in the spectrum.

It is notable that with \menate included the residual data--simulation discrepancy amounts to an apparent normalization that is nearly uniform across the non-energy variables (i.e., \crefrange{fig:after_cuts_geant_b}{fig:after_cuts_geant_d} versus \crefrange{fig:after_cuts_menate_b}{fig:after_cuts_menate_d}). In the simulation chain, the (anti) neutrino interaction model is responsible for producing the primary final-state neutrons. These models are subject to significant uncertainties related both to the details of the primary neutrino interaction and to the final-state interactions that occur before the particles are handed over to Geant4 for propagation through the detector. In turn, there is significant variance in the simulated final-state neutron multiplicity \cite{MINERvA:2019,SK:2025,T2K:2025}. The consistency of the residual disagreement with an overall normalization factor implicates the neutrino interaction model as a potential source of the remaining excess of neutron candidates in the simulation.

\begin{figure}[t]
    \centering
    \subfloat[]{\includegraphics[width=.49\textwidth]
    {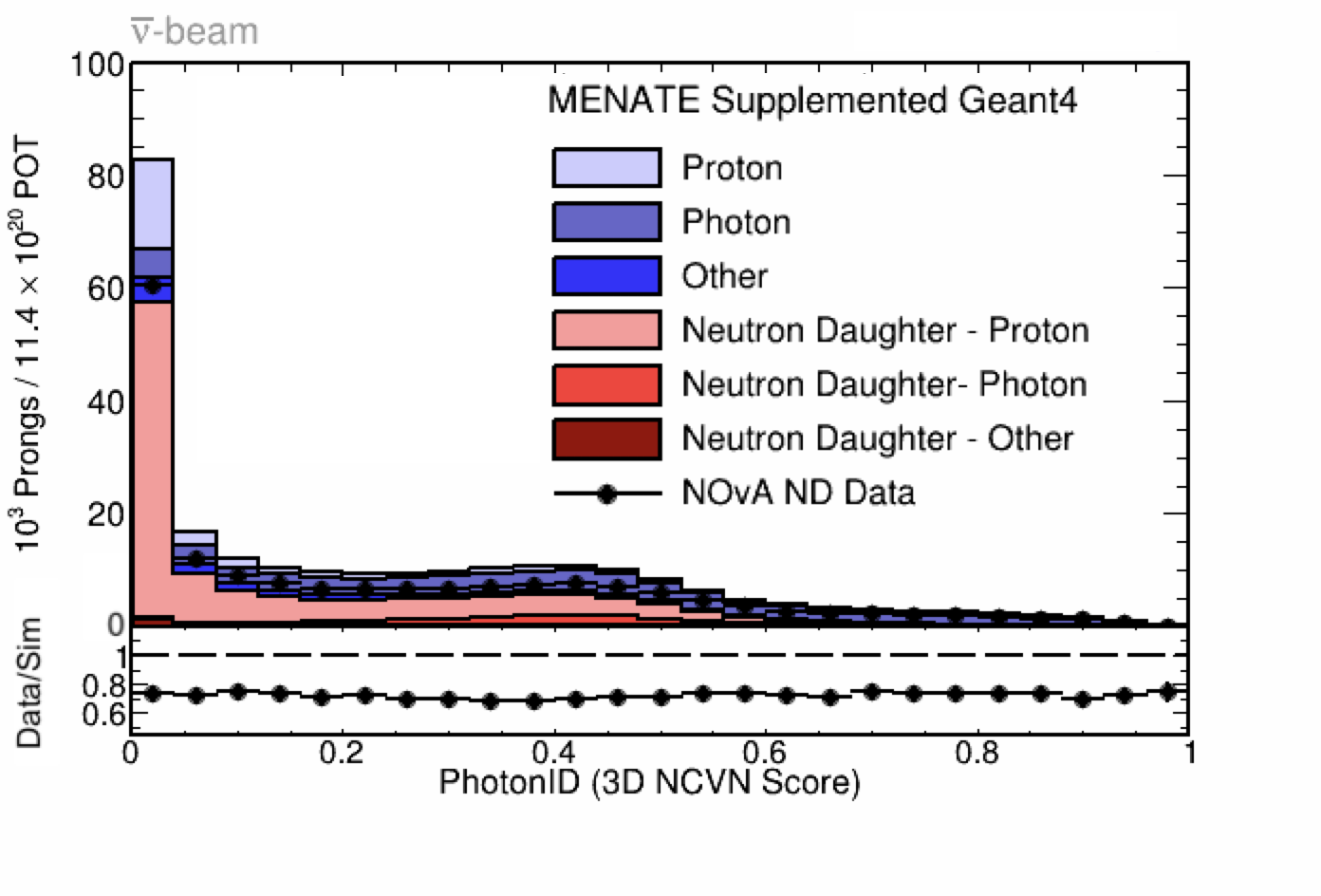}\label{fig:3dcvn_scores_menate}}
    \subfloat[]{\includegraphics[width=.49\textwidth]
    {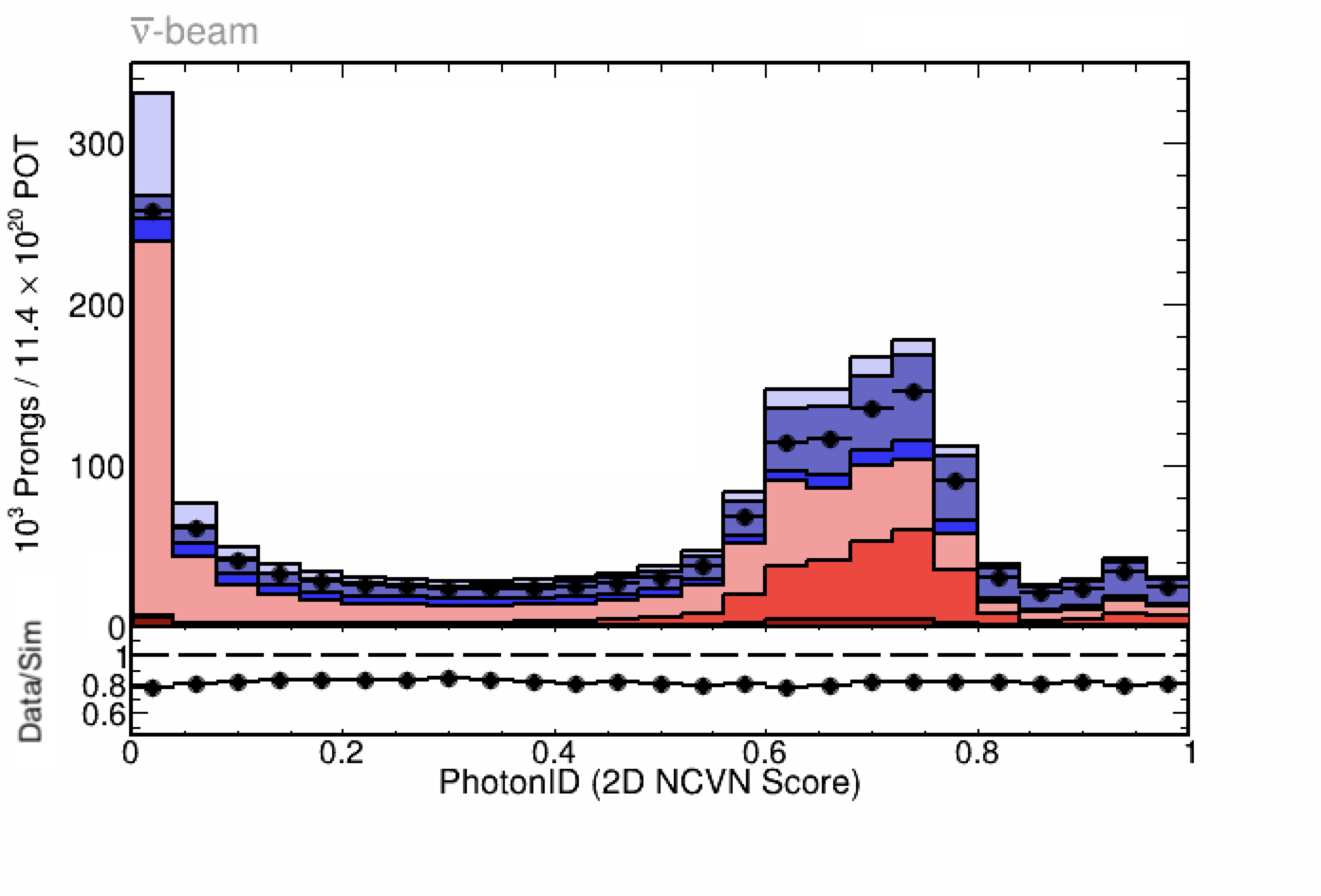}\label{fig:2dcvn_scores_menate}}
    \caption{Photon identification scores from the neutron CNNs applied to neutron candidate prongs in data and the \menate-supplemented Geant4 simulation. (a) Three-dimensional prongs and (b) two-dimensional prongs.}
    \label{fig:cvn_scores_menate}
\end{figure}

The impact of \menate on the secondary-prong content of the neutron sample is even more visible in the particle-identification scores from the neutron CNNs. A comparison of Figure~\ref{fig:cvn_scores_menate} with Figure~\ref{fig:cvn_scores_geant} shows a dramatic reduction in the photon peak at CNN scores between 0.6 and 0.8. Recalling the source of secondary photons from neutron interactions, the evaporation model used by the Bertini cascade in Geant4 is concluded to be a significant contributor to the observed simulation excess.

\section{Summary and discussion}
\label{sec:summary}

Accounting for the portion of an incoming neutrino's energy that is carried away by neutrons is one of the largest sources of systematic uncertainty in many neutrino analyses and will likely play an important role in future experiments as well~\cite{NOvA:2021nfi, MINERvA:2023avz, MicroBooNE:2022cls, DUNE:TDR}. 
An algorithm has been developed that selects neutron-associated energy depositions (referred to as prongs) within the \nova ND using simple cuts on two low-level reconstructed variables motivated by the known characteristics of neutrons. The neutron-candidate selection identifies 73\% of all prongs produced by a secondary particle from the interaction of a primary neutron. Applying the selection to both \nova's default Geant4 simulation and the data reveals a large excess of neutron-prong candidates in the simulation. Evidence is seen for a link between this excess and prongs produced by neutron-induced photon secondaries. 

The \menate interaction model as a supplement to the Geant4 intranuclear cascade for neutron-on-carbon interactions between \SIrange{20}{100}{\mega\eV} has been shown to provide better agreement with \nova ND data. The most significant difference between \menate and Geant4 lies in the inelastic neutron modeling details. The total inelastic cross sections are very similar, but \menate uses individual final-state particle cross sections compiled directly from scattering measurements. Geant4 uses a total inelastic neutron cross section and statistically determines the outgoing particles in a way that can produce final states not observed in the data. The most significant physical change provided by \menate is a reduction in the number of visible photon secondaries.   

\begin{figure}[t]
    \centering
 \subfloat[]{\includegraphics[width=.6\textwidth]{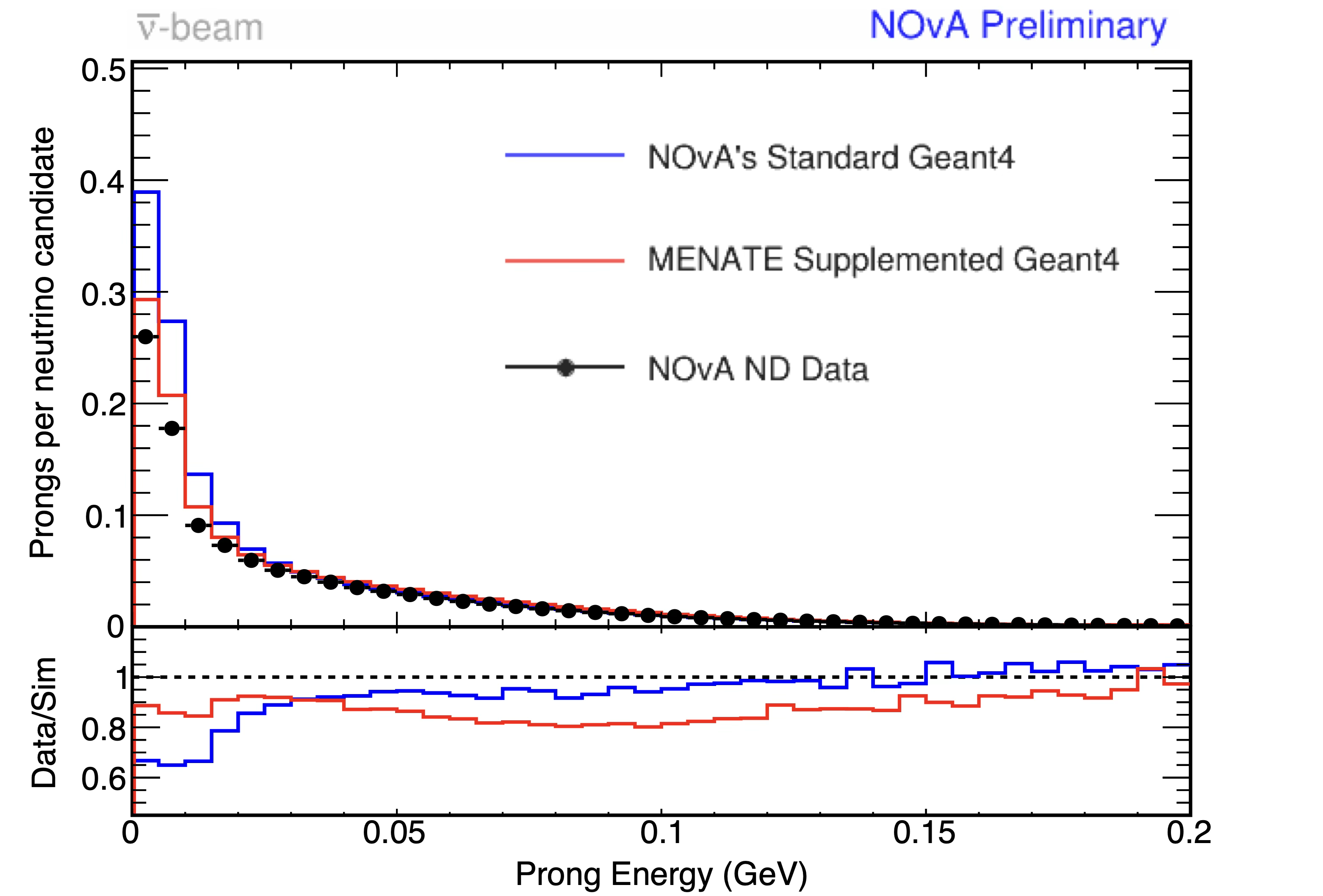}}    
    \caption{Comparison of calorimetric energy for neutron-candidate prongs selected in data (black points), Geant4-only (blue), and MENATE-supplemented Geant4 (red). All samples are normalized by the number of selected neutrino candidates in each.}
    \label{fig:prongE_comps}
\end{figure}

A direct comparison of data to the simulations is shown in Figure~\ref{fig:prongE_comps}, where each sample is normalized by the number of selected neutrino candidates. This accounts for discrepancies arising from the beam flux and neutrino interaction model uncertainties that impact the neutrino event rate, and reduces the over-simulation by about 5\%. The remaining  excess of simulated neutron-candidate prongs as a function of calorimetric energy decreases from a maximum of $\sim$35\% in the most populated region down to $\sim$15\%. While some shape differences remain in the calorimetric energy distribution, the over-simulation in \menate-supplemented sample is much more uniform compared to the base Geant4 simulation, and is very flat across the distributions of other reconstructed variables from Figure~\ref{fig:after_cuts_menate}. This supports the conclusion that \menate provides significantly better agreement with \nova ND data and that the remaining oversimulation is likely due to mismodeling of neutron production in the primary neutrino interaction model.

Using the \menate-supplemented simulation as the basis for the neutron systematic uncertainty, rather than an ad hoc shift in neutron-induced prong energies, reduces the impact of the neutron systematic by about a factor of two when comparing the most recent \nova three-flavor oscillation analysis~\cite{NOvA:2025hbg} with the previous measurement~\cite{NOvA:2021nfi}. Based on these findings, \nova has adopted \menate, in addition to the Geant4 configuration described above, as the primary neutron transport and interaction model for future simulation productions. 

\acknowledgments
\label{z_ackn}
This document was prepared by the NOvA collaboration using the resources of the Fermi National Accelerator Laboratory (Fermilab), a U.S. Department of Energy, Office of Science, HEP User Facility. Fermilab is managed by Fermi Forward Discovery Group, LLC, acting under Contract No. 89243024CSC000002.  This work was supported by the U.S. Department of Energy; the U.S. National Science Foundation; the Department of Science and Technology, India; the European Research Council; the MSMT CR, GA UK, Czech Republic; the RAS, the Ministry of Science and Higher Education, and RFBR, Russia; CNPq and FAPEG, Brazil; UKRI, STFC and the Royal Society, United Kingdom; and the state and University of Minnesota.  We are grateful for the contributions of the staffs of the University of Minnesota at the Ash River Laboratory, and of Fermilab. For the purpose of open access, the author has applied a Creative Commons Attribution (CC BY) license to any Author Accepted Manuscript version arising.

\printauthorlist

\bibliographystyle{JHEP}
\bibliography{refs.bib}

@article{T2K:2025yoy,
    author = "Abe, K. and others",
    collaboration = "T2K",
    title = "{Results from the T2K Experiment on Neutrino Mixing Including a New Far Detector {\ensuremath{\mu}}-like Sample}",
    eprint = "2506.05889",
    archivePrefix = "arXiv",
    primaryClass = "hep-ex",
    doi = "10.1103/gh5j-5cwv",
    journal = "Phys. Rev. Lett.",
    volume = "135",
    number = "26",
    pages = "261801",
    year = "2025"
}

@article{NOvA:2025hbg,
    author = "Abubakar, S. and others",
    collaboration = "NOvA",
    title = "{Precision Measurement of Neutrino Oscillation Parameters with 10 Years of Data from the NOvA Experiment}",
    eprint = "2509.04361",
    archivePrefix = "arXiv",
    primaryClass = "hep-ex",
    reportNumber = "FERMILAB-PUB-25-0619-PPD",
    doi = "10.1103/x53y-2b86",
    journal = "Phys. Rev. Lett.",
    volume = "136",
    number = "1",
    pages = "011802",
    year = "2026"
}

@article{NOvAT2K:2025wet,
    author = "Abubakar, S. and others",
    collaboration = "T2K, NOvA",
    title = "{Joint neutrino oscillation analysis from the T2K and NOvA experiments}",
    eprint = "2510.19888",
    archivePrefix = "arXiv",
    primaryClass = "hep-ex",
    reportNumber = "FERMILAB-PUB-25-0132-PPD",
    doi = "10.1038/s41586-025-09599-3",
    journal = "Nature",
    volume = "646",
    number = "8086",
    pages = "818--824",
    year = "2025"
}

@article{DUNE2022,
    title = {Deep underground neutrino experiment: DUNE}, 
    author = {Falcone, A. and others},
    year = {2022},
    journal = "Nuclear Instruments and Methods in Physics Research Section A: Accelerators, Spectrometers, Detectors and Associated Equipment",
    volume = "1041",
    doi = "10.1016/j.nima.2022.167217",
    eprint = "2103.13910",
    archivePrefix = "arXiv",
}

@techreport{Blokhin:2016kje,
  author = {Blokhin, A. I. and others},
  title = {New version of neutron evaluated data library {BROND-3.1}},
  year = {2016},
  type = {{}},
  institution = {{}},
  address = {{}},
  number = {Yad. Reak. Konst. No. 2, p. 62},
  url = "https://vant.ippe.ru/en/brond-3-1"
}

@article{Zhigang:2020kel,
    author = "{Zhigang Ge} and others",
    title = "{CENDL-3.2: The new version of Chinese general purpose evaluated nuclear data library}",
    doi = "10.1051/epjconf/202023909001",
    journal = "EPJ Web Conf.",
    volume = "239",
    pages = "09001",
    year = "2020"
}

@article{Brown:2018jhj,
    author = "Brown, D. A. and others",
    title = "{ENDF/B-VIII.0: The 8th Major Release of the Nuclear Reaction Data Library with CIELO-project Cross Sections, New Standards and Thermal Scattering Data}",
    doi = "10.1016/j.nds.2018.02.001",
    journal = "Nucl. Data Sheets",
    volume = "148",
    pages = "1--142",
    year = "2018"
}

@article{Plompen:2020yhs,
    author = "Plompen, A.J.M. and others",
    title = "{The joint evaluated fission and fustion nuclear data library, JEFF-3.3}",
    doi = {10.1140/epja/s10050-020-00141-9},
    journal = {Eur. Phys. J. A.},
    volume = "56",
    pages = "181",
    year = "2020"
}

@article{Osamu:2023ouf,
    author = {Osamu Iwamoto and others},
    title = {Japanese evaluated nuclear data library version 5: JENDL-5},
    journal = {Journal of Nuclear Science and Technology},
    volume = {60},
    number = {1},
    pages = {1-60},
    year = {2023},
    doi = {10.1080/00223131.2022.2141903}
}

@article{Koning:2019fks,
    title = {TENDL: Complete Nuclear Data Library for Innovative Nuclear Science and Technology},
    journal = {Nuclear Data Sheets},
    volume = {155},
    pages = {1-55},
    year = {2019},
    doi = {https://doi.org/10.1016/j.nds.2019.01.002},
    url = {https://www.sciencedirectx.com/science/article/pii/S009037521930002X},
    author = {A.J. Koning and others},
}

@article{DelGuerra:1975stf,
    author = "Del Guerra, A.",
    title = "{A Compilation of n p and n c Cross-Sections and their Use in a Monte Carlo Program to Calculate the Neutron Detection Efficiency in Plastic Scintillator in the Energy Range 1-MeV to 300-MeV}",
    reportNumber = "DL-P-245",
    doi = "10.1016/0029-554X(76)90181-6",
    journal = "Nucl. Instrum. Meth.",
    volume = "135",
    pages = "337",
    year = "1976"
}

@article{Mufson:2015kga,
    author = "Mufson, S. and others",
    title = "{Liquid scintillator production for the NOvA experiment}",
    eprint = "1504.04035",
    archivePrefix = "arXiv",
    primaryClass = "physics.ins-det",
    reportNumber = "FERMILAB-PUB-15-048-ND-PPD",
    doi = "10.1016/j.nima.2015.07.026",
    journal = "Nucl. Instrum. Meth. A",
    volume = "799",
    pages = "1--9",
    year = "2015"
}

@article{NOvA:2021nfi,
    author = "Acero, M. A. and others",
    title = "{Improved measurement of neutrino oscillation parameters by the NOvA experiment}",
    eprint = "2108.08219",
    archivePrefix = "arXiv",
    primaryClass = "hep-ex",
    reportNumber = "FERMILAB-PUB-21-373-ND",
    doi = "10.1103/PhysRevD.106.032004",
    journal = "Phys. Rev. D",
    volume = "106",
    number = "3",
    pages = "032004",
    year = "2022"
}

@article{MINERvA:2023avz,
    author = "Cai, T. and others",    
    title = "{Measurement of the axial vector form factor from antineutrino\textendash{}proton scattering}",
    reportNumber = "FERMILAB-PUB-23-033-CSAID-ND-QIS",
    doi = "10.1038/s41586-022-05478-3",
    journal = "Nature",
    volume = "614",
    number = "7946",
    pages = "48--53",
    year = "2023"
}

@article{MicroBooNE:2022cls,
    author = "Abratenko, P. and others",
    title = "{First Measurement of Quasielastic \ensuremath{\Lambda} Baryon Production in Muon Antineutrino Interactions in the MicroBooNE Detector}",
    eprint = "2212.07888",
    archivePrefix = "arXiv",
    primaryClass = "hep-ex",
    reportNumber = "FERMILAB-PUB-22-925-ND",
    doi = "10.1103/PhysRevLett.130.231802",
    journal = "Phys. Rev. Lett.",
    volume = "130",
    number = "23",
    pages = "231802",
    year = "2023"
}

@article{DUNE:TDR,
    author = "Abi, Babak and others",
    collaboration = "DUNE",
    title = "{Deep Underground Neutrino Experiment (DUNE), Far Detector Technical Design Report, Volume I Introduction to DUNE}",
    eprint = "2002.02967",
    archivePrefix = "arXiv",
    primaryClass = "physics.ins-det",
    reportNumber = "FERMILAB-PUB-20-024-ND, FERMILAB-DESIGN-2020-01",
    doi = "10.1088/1748-0221/15/08/T08008",
    journal = "JINST",
    volume = "15",
    number = "08",
    pages = "T08008",
    year = "2020"
}

@article{menateCreation,
    author = "Desesquelles, P. and others",
    title = "{Cross talk and diaphony in neutron detectors}",
    journal = "Nucl. Instrum. Methods Phys. Res. A",
    year = "1991",
    volume = "307",
    issue = "2-3",
    pages = "366--373",
    doi = "10.1016/0168-9002(91)90206-6"
}

@article{Agostinelli:2003yb,
    author    = "Agostinelli, S. and others",
    title     = "{Geant4: A simulation toolkit}",
    journal   = "Nucl. Instrum. Meth.",
    volume    = "A506",
    year      = "2003",
    pages     = "250-303",
    doi       = "10.1016/S0168-9002(03)01368-8"
}

@article{NumI_2016,
    title="{The NuMI neutrino beam}",
    volume="806",
    ISSN="0168-9002",
    DOI="10.1016/j.nima.2015.08.063",
    journal="Nuclear Instruments and Methods in Physics Research Section A: Accelerators, Spectrometers, Detectors and Associated Equipment",
    publisher="Elsevier BV",
    author="Adamson, P. and others",
    year="2016",   
    pages="279–306" ,
    eprint = "1507.06690",
    archivePrefix = "arXiv",
    primaryClass = "hep-ex",
}

@article{Andreopoulos:2009rq,
    author = "Andreopoulos, C. and others",
    title = "{The GENIE Neutrino Monte Carlo Generator}",
    eprint = "0905.2517",
    archivePrefix = "arXiv",
    primaryClass = "hep-ph",
    reportNumber = "FERMILAB-PUB-09-418-CD",
    doi = "10.1016/j.nima.2009.12.009",
    journal = "Nucl. Instrum. Meth. A",
    volume = "614",
    pages = "87--104",
    year = "2010"
}

@book{Andreopoulos:2015wxa,
    author = "Andreopoulos, C. and others",
    title = "{The GENIE Neutrino Monte Carlo Generator: Physics and User Manual}",
    eprint = "1510.05494",
    archivePrefix = "arXiv",
    primaryClass = "hep-ph",
    reportNumber = "FERMILAB-FN-1004-CD",
    month = "10",
    year = "2015",
    publisher = {{}},
}

@techreport{G4PhysMan:2017aa,
  author = {{GEANT4 Collaboration}},
  title = "{GEANT4 Physics Reference Manual}",
  year = {2017},
  number = {Release 10.4},
  type = {{}},
  institution = {{}},
  address = {{}},
  url = "https://geant4-userdoc.web.cern.ch/UsersGuides/PhysicsReferenceManual/BackupVersions/V10.4/fo/PhysicsReferenceManual.pdf"
}

@article{Dostrovsky:1959zz,
    author = "Dostrovsky, I. and others",
    title = "{Monte Carlo Calculations of Nuclear Evaporation Processes. 3. Applications to Low-Energy Reactions}",
    doi = "10.1103/PhysRev.116.683",
    journal = "Phys. Rev.",
    volume = "116",
    pages = "683--702",
    year = "1959"
}

@article{PhysRev.118.791,
    title = {Monte Carlo Calculations of Nuclear Evaporation Processes. V. Emission of Particles Heavier Than {He4}},
    author = {Dostrovsky, I. and others},
    journal = {Phys. Rev.},
    volume = {118},
    issue = {3},
    pages = {791--793},
    year = {1960},
    month = {May},
    doi = {10.1103/PhysRev.118.791},
    url = {https://link.aps.org/doi/10.1103/PhysRev.118.791}
}

@article{Chadwick:2011endf,
    author = "Chadwick, M. B. and others",
    title = "{ENDF/B-VII.1 Nuclear Data for Science and Technology: Cross Sections, Covariances, Fission Product Yields and Decay Data}",
    doi = "10.1016/j.nds.2011.11.002",
    journal = "Nucl. Data Sheets",
    volume = "112",
    pages = "2887--2996",
    year = "2011"
}

@article{ref:mona, 
    author = "Kohley, Z. and others",
    title = "{Modeling interactions of intermediate-energy neutrons in a plastic scintillator array with Geant4}",
    journal = "Nucl. Instrum. Methods Phys. Res. A",
    volume = "682",
    pages = "59--65",
    year = "2012",
    doi = {10.1016/j.nima.2012.04.060}
}

@article{Cecil:1979,
    title = {Improved predections of neutron detection efficiency for hydrocarbon scintillators from 1 {MeV} to about 300 {MeV}},
    journal = {Nuclear Instruments and Methods},
    volume = {161},
    number = {3},
    pages = {439-447},
    year = {1979},
    issn = {0029-554X},
    doi = {https://doi.org/10.1016/0029-554X(79)90417-8},
    url = {https://www.sciencedirect.com/science/article/pii/0029554X79904178},
    author = {Cecil, R.A. and others}
}

@article{Guerra:1976,
    title = {A compilation of n-p and n-C cross sections and their use in a Monte Carlo program to calculate the neutron detection efficiency in plastic scintillator in the energy range 1–300 {MeV}},
    journal = {Nuclear Instruments and Methods},
    volume = {135},
    number = {2},
    pages = {337-352},
    year = {1976},
    issn = {0029-554X},
    doi = {https://doi.org/10.1016/0029-554X(76)90181-6},
    url = {https://www.sciencedirect.com/science/article/pii/0029554X76901816},
    author = {{Del Guerra}, A.}
}

@article{NPTool:2016aa,
    title = {NPTool: a simulation and analysis framework for low-energy nuclear physics experiments},
    author = {Matta, A and others},
    issn = {0954-3899},
    journal = {Journal of Physics G-Nuclear and Particle Physics},
    number = {4},
    publisher = {IOP Publishing LTD},
    volume = {43},
    year = {2016},
    doi = {10.1088/0954-3899/43/4/045113}
}

@article{NOvA:2019nfi,
    author = "Acero, M. A. and others",
    title = "{First Measurement of Neutrino Oscillation Parameters using Neutrinos and Antineutrinos by NOvA}",
    doi = "10.1103/PhysRevLett.123.151803",
    journal = "Phys. Rev. Lett.",
    volume = "123",
    issue = "15",
    pages = "151803",
    numpages = "8",
    year = "2019",
    month = "Oct",
    publisher = "American Physical Society",
    url = "https://link.aps.org/doi/10.1103/PhysRevLett.123.151803",
    eprint = "1906.04907",
    archivePrefix = "arXiv",
}

@article{sandler2019mobilenetv2,
    author = {Sandler, Mark and others},
    journal = { 2018 IEEE/CVF Conference on Computer Vision and Pattern Recognition (CVPR) },
    title = {{ MobileNetV2: Inverted Residuals and Linear Bottlenecks }},
    year = {2018},
    volume = {},
    ISSN = {},
    pages = {4510-4520},
    doi = {10.1109/CVPR.2018.00474},
    url = {https://doi.ieeecomputersociety.org/10.1109/CVPR.2018.00474},
    publisher = {IEEE Computer Society},
    address = {Los Alamitos, CA, USA},
    month =Jun,
    eprint="1801.04381",
    archivePrefix="arXiv",
}

@article{Psihas:2019ksa,
    author = "Psihas, F. and others",
    title = {Context-Enriched Identification of Particles with a Convolutional Network for Neutrino Events},
    eprint = "1906.00713",
    archivePrefix = "arXiv",
    primaryClass = "physics.ins-det",
    reportNumber = "FERMILAB-PUB-19-258-PPD",
    doi = "10.1103/PhysRevD.100.073005",
    journal = "Phys. Rev. D",
    volume = "100",
    number = "7",
    pages = "073005",
    year = "2019"
}

@article{MINERvA:2019,
    title = {Neutron measurements from antineutrino hydrocarbon reactions},
    author = {Elkins, M. and others},  
    eprint = "1901.04892",
    archivePrefix = "arXiv",
    journal = {Phys. Rev. D},
    volume = {100},
    issue = {5},
    pages = {052002},
    numpages = {20},
    year = {2019},
    month = {Sep},
    publisher = {American Physical Society},
    doi = {10.1103/PhysRevD.100.052002},
    url = {https://link.aps.org/doi/10.1103/PhysRevD.100.052002}
}

@article{SK:2025,
    title = {Measurement of neutron production in atmospheric neutrino interactions at Super-Kamiokande},
    author = {Han, S. and others},
    eprint = "2505.04409",
    archivePrefix = "arXiv",
    journal = {Phys. Rev. D},
    volume = {112},
    issue = {1},
    pages = {012004},
    numpages = {24},
    year = {2025},
    month = {Jul},
    publisher = {American Physical Society},
    doi = {10.1103/4d71-d69k},
    url = {https://link.aps.org/doi/10.1103/4d71-d69k}
}

@article{T2K:2025,
    title = {First measurement of neutron capture multiplicity in neutrino-oxygen neutral-current quasielasticlike interactions using an accelerator neutrino beam},
    author = {Abe, K. and others},
    eprint = "2505.22547",
    archivePrefix = "arXiv",
    journal = {Phys. Rev. D},
    volume = {112},
    issue = {3},
    pages = {032003},
    numpages = {26},
    year = {2025},
    month = {Aug},
    publisher = {American Physical Society},
    doi = {10.1103/qh28-4znk},
    url = {https://link.aps.org/doi/10.1103/qh28-4znk}
}

\end{document}